\RequirePackage{etoolbox}
\documentclass[ba]{imsart}
\firstpage{1}
\lastpage{1}

\usepackage{amsthm}
\usepackage{amsmath}
\usepackage{natbib}
\usepackage[algo2e]{algorithm2e}
\usepackage{algorithmic}  
\usepackage{algorithm}
\usepackage{mathtools}
\usepackage[colorlinks,citecolor=blue,urlcolor=blue,filecolor=blue,backref=page]{hyperref}
\usepackage{graphicx}
\usepackage{subcaption}

\def\spacingset#1{\renewcommand{\baselinestretch}%
	{#1}\small\normalsize} \spacingset{1}
\usepackage{fmtcount}
\startlocaldefs
\numberwithin{equation}{section}
\theoremstyle{plain}

\endlocaldefs
\DeclarePairedDelimiter\abs{\lvert}{\rvert}
\begin{document}
	
	\begin{frontmatter}
		\title{The Hyperedge Event Model}
		\runtitle{The Hyperedge Event Model}
		
		\begin{aug}
			\author{\fnms{Bomin} \snm{Kim}\thanksref{addr1,t1,t2,m1}\ead[label=e1]{bzk147@psu.edu}},
			\author{\fnms{Aaron} \snm{Schein}\thanksref{addr2,t3,m1,m2}\ead[label=e2]{aschein@cs.umass.edu}},
			\author{\fnms{Bruce} \snm{A. Desmarais}\thanksref{addr3,t3,m1,m2}\ead[label=e3]{bdesmarais@psu.edu}},
			\and
			\author{\fnms{Hanna} \snm{Wallach}\thanksref{addr4,t1,m2}
				\ead[label=e4]{hanna@dirichlet.net}
			}
			\runauthor{B. Kim et al.}
			
			\address[addr1]{Department of Statistics, Pennsylvania State University
				\printead{e1} 
			}
			\address[addr2]{College of Information and Computer Sciences, UMass Amherst
				\printead{e2} 
			}
			\address[addr3]{Department of Political Science, Pennsylvania State University
				\printead{e3} 
			}
			\address[addr4]{Microsoft Research NYC
				\printead{e4}
			}
			
			
		\end{aug}
		
		\begin{abstract}
			We introduce the hyperedge event model (HEM)---a generative model for events that can be represented as directed edges with one sender and one or more receivers or one receiver and one or more senders. We
			integrate a dynamic version of the exponential random graph model (ERGM) of edge structure with a survival model for event timing to jointly understand who interacts with whom, and when. The HEM offers three innovations with respect to the literature---first, it extends a growing class of dynamic network models to model hyperedges. The current state-of-the-art approach to dealing with hyperedges is to inappropriately break them into separate edges/events. Second, our model involves a novel receiver selection distribution that is based on established edge formation models, but assures non-empty receiver lists. Third, the HEM integrates separate, but interacting, equations governing edge formation and event timing. We use the HEM to analyze emails sent among department managers in Montgomery County government in North Carolina. Our application demonstrates that the model is effective at predicting and explaining time-stamped network data involving edges with multiple receivers. We present an out-of-sample prediction experiment to illustrate how researchers can select between different specifications of the model. 
		\end{abstract}
		
		
		\begin{keyword}
			\kwd{dynamic network model} 
			\kwd{hyperedge}
			\kwd{continuous time model}
			\kwd{email data analysis}
		\end{keyword}
		
	\end{frontmatter}
	
	\section{Introduction}\label{sec:introduction}
	
	Processes that arise as time-stamped directed interactions are common in the social, natural, and phyiscal sciences. The data produced by such processes can be represented as dynamic directed networks---an object that has given rise to the development of several statistical model families. For example, stochastic actor-oriented models (SAOMs) \citep{snijders1996stochastic,snijders2007modeling} characterize that network evolutions occur as the senders decide to create or remove an edge from the existing network, one edge at a time. Event-based network models \citep{Butts2008,Vu2011,hunter2011dynamic,PerryWolfe2012} provide a general framework for modeling the realization of edges that occur as instances in continuous time streams of events. This family of models is flexible enough to account for the many ways in which past network structures beget future ones---e.g., if node $i$ directed a tie to node $j$ recently, then node $j$ will direct one to node $i$ in the near future---and useful for understanding the traits and behaviours that are predictive of interactions. 
	
	A major limitation of existing dynamic network models is that they apply to edges with one sender and one receiver. Dynamic network data often naturally arise as ``hyperedges'' \citep{karypis1999multilevel,ghoshal2009random,zlatic2009hypergraph,zhang2010hypergraph} that include one sender and multiple receivers or one receiver and multiple senders. For example, in email networks \citep{newman2002email}, each email encodes a hyperedge from one sender to one or more receviers. Networks formed between neurons via axons and dendrites involve hyperedges with one sender and multiple receivers (axons) or one receiver and multiple senders (dendrites) \citep{partzsch2012developing}. Networks formed through the cosponsorhip of legislative bills \citep{fowler2006legislative} involve hyperedges with multiple senders (cosponsors) and one receiver (sponsor). Economic sanctions between countries \citep{cranmer2014reciprocity} induce networks with hyperedges between multiple sending countries and one target country. Existing models require researchers to alter hyperedge data to fit with the pairwise edge structure of the model. For instance, \cite{PerryWolfe2012} treat multicast interactions---one type of directed hyperedge which involves one sender and one or more receivers---via duplication (i.e., obtain pairwise interactions from the original multicast), to construct an approximate likelihood function in their inferential framework for model parameters. Similarly, \cite{fan2009learning} duplicate emails sent from one sender to one or more receivers and randomly jitter the sent times in order to avoid violating the assumption that two events cannot occur at the exact same time. 
	
	We develop a statistical dynamic network model, which we term the hyperedge event model (HEM), that integrates the two components that govern hyperedge event formation: (1) the formation of the vertices that are incident to the hyperedge, and (2) the timing of the hyperedge event. In what follows, we define the HEM's generative process for hyperedge event data (Section \ref{sec:generative process}), derive the conditional posteriors for Bayesian inference, and present tests of our software implementation (Section \ref{sec:inference}). We then demonstrate our model's applicability in a case study (Section \ref{sec:Emails}) where we analyze a corpus of internal county government emails and illustrate how to perform model selection, posterior predictive checks, and exploratory analysis using our model. We conclude in Section \ref{sec:conclusion}.~\looseness=-1
	
	\section{The hyperedge event model}\label{sec:generative process}
	

	The hyperedge event model (HEM) specifies a generative process for $E$ unique hyperedge events that occur between $A$ nodes. A single hyperedge event is indexed by $e \in [E]$---where $[E]$ denotes a categorical set with $E$ levels $[E] = \{1,\ldots,E\}$---and consists of three components: the sender $s_e \in [A]$, an indicator vector of receivers $\boldsymbol{r}_e$---where $r_{ej}=1$ if $j \in [A]$ is a receiver of hyperedge event $e$ and 0 otherwise---and the timestamp $t_e \in (0, \infty)$. For simplicity, we assume that events are ordered by time such that $t_e \leq t_{e+1}$. While the model can be applied to two types of hyperedge events---events involving (1) one sender and one or more receivers, and (2) one or more senders and one receiver---here we only present the generative process for those involving one sender and one or more receivers (i.e., multicast). One notable feature of our generative process is that we draw auxiliary variables that serve as candidate data. Data is generated from the HEM through a sampling process applied to the auxiliary variables. The auxiliary variables drawn for event $e$ include, for each possible sender $i \in [A]$, a time increment from event $e-1$ at which sender $i$ would create event $e$, and an $A-1$ length vector indicating which nodes would be the receivers of event $e$ if it were directed by sender $i$.  The data generated for event $e$ under the HEM corresponds to the sender that would create event $e$ the soonest---at the smallest time increment from event $e-1$. The receivers of event $e$ generated under the HEM correspond to those receivers toward which the sender with the minimum time increment would have directed event $e$. We explain these steps in more detail below. For hyperedge events that involve one receiver and one or more senders, we treat $s_e$ to be an indicator vector of senders $\boldsymbol{s}_e$ and $r_e$ to be the single receiver, and then follow the alternative generative process provided in Appendix A, which we derive as a simple reversal of the process used for multicasts (i.e., one sender and multiple receivers).~\looseness=-1 
	
	\subsection{Candidate receivers}\label{subsec: Tie}
	For every possible sender--receiver pair $(i,j)$ where $i \!\neq\! j$, we define the ``receiver intensity" as a linear combination of statistics relevant to the receiver selection process:
	\begin{equation}
		\lambda_{iej} = {\boldsymbol{b}}^{\top}\boldsymbol{x}_{iej},
	\end{equation}
	where $\boldsymbol{b}$ is a $P$-dimensional vector of coefficients and $\boldsymbol{x}_{iej}$ is a set of receiver selection features. As we show below, this intensity contributes to the probability that $i$ directs event $e$ towards receiver $j$. The features $\boldsymbol{x}_{iej}$ can capture common network processes like popularity, reciprocity, and transitivity, as well as the effects of attributes of the sender and receivers (e.g., their gender), or attributes of sender--receiver pairs (e.g., whether the sender is a supervisor of the receiver's). We also include a normally distributed intercept term to account for the average (or baseline) number of receivers: $\boldsymbol{b} \sim N(\boldsymbol{\mu}_b, \Sigma_b)$.~\looseness=-1
	
	The HEM assumes that the sender of each hyperedge event is the sender that would initiate their respective event with the greatest urgency (i.e., the earliest timestamp). Our model thus assumes that for every event $e$, every possible sender $i$ generates a candidate receiver set that would be the receiver set of event $e$ if sender $i$ were the sender. For an event $e$, we first define an $A\times A$ matrix $\boldsymbol{u}_e$ where the $i^{\textrm{th}}$ row denotes sender $i$'s receiver vector of zeros and 1's---i.e., 1's indicate the nodes to which $i$ intends to direct event $e$. We then assume that each receiver vector $\boldsymbol{u}_{ie}$ comes from a modification of the multivariate Bernoulli (MB) distribution \citep{dai2013multivariate}---a model that has been used to model graphs in which the state of each edge indicator is drawn independently from an edge-specific Bernoulli distribution. In order to avoid drawing hyperedge events with no receivers, we define a probability measure ``MB$_{G}$" motivated by the Gibbs measure \citep{fellows2017removing}. The probability measure we define amounts to a non-empty Gibbs measure, in which the all-zero vector is excluded from the support of the multivariate Bernoulli distribution. As a result, this measure helps us to (1) allow a sender to select multiple receivers for a single event, (2) force the sender to select at least one receiver, and (3) ensure a tractable normalizing constant for the receiver selection distribution. To be specific, we draw a binary vector $\boldsymbol{u}_{ie}= (u_{ie1},
	\ldots, u_{ieA})$ 
	\begin{equation} \boldsymbol{u}_{ie}  \sim
		\mbox{MB}_{G}(\boldsymbol{\lambda}_{ie}),
	\end{equation}
	where $\boldsymbol{\lambda}_{ie}= (\lambda_{ie1},
	\ldots, \lambda_{ieA})$. In particular, we define $\mbox{MB}_{G}(\boldsymbol{\lambda}_{ie})$ as
	\begin{equation}
		\begin{aligned}
			&\Pr(\boldsymbol{u}_{ie}|\,\boldsymbol{b}, \boldsymbol{x}_{ie}) = \frac{1}{Z(\boldsymbol{\lambda}_{ie})}\exp\Big(\mbox{log}\big(\text{I}( \lVert \boldsymbol{u}_{ie}\rVert_1 > 0 )\big) + \sum_{j\neq i} \lambda_{iej}u_{iej}\Big) ,
		\end{aligned}
		\label{eqn:Gibbs}
	\end{equation}
	where $Z(\boldsymbol{\lambda}_{ie})$ is the normalizing constant, $\lVert \cdot \rVert_1$ is the $\ell_1$-norm, and the log-indicator term $\mbox{log}\big(\text{I}( \lVert \boldsymbol{u}_{ie}\rVert_1 > 0 )\big)$ ensures that empty receiver sets are excluded from the distribution's support. These modeling assumptions facilitate efficient posterior inference since we can derive a closed form expression for the normalizing constant---i.e., $Z(\boldsymbol{\lambda}_{ie})= \prod_{j\neq i} \big(\mbox{exp}(\lambda_{iej}) + 1\big)-1$---and thus do not need to perform brute-force summation over the support of $\boldsymbol{u}_{ie} \in [0,1]^A$. We provide detailed derivation steps for the normalizing constant $Z(\boldsymbol{\lambda}_{ie})$ in Appendix B. 

	
	\subsection{Candidate timestamps}\label{subsec:Time}
	To generate timing for each event, the HEM first draws a candidate timestamp at which the event would be created given the candidate sender and receiver combinations. The timing rate for sender $i$ and event $e$ is
	\begin{equation}
		\mu_{ie} = g^{-1}(\boldsymbol{\eta}^\top \boldsymbol{y}_{ie}),
	\end{equation}
	where $\boldsymbol{\eta}$ is a $Q$-dimensional vector of coefficients with a Normal prior $\boldsymbol{\eta} \sim N(\boldsymbol{\mu}_\eta,\Sigma_\eta)$, $\boldsymbol{y}_{ad}$ is a set of event timing features---covariates that could affect timestamps of events, and $g(\cdot)$ is the appropriate link function such as identity, log, or inverse. 
	
	In modeling ``when," we do not directly model the timestamp $t_e$. Instead, we assume that each sender's ``time increment"---i.e., waiting time to next event since $t_{e-1}$---is drawn from a specific exponential family distribution. We define the time increment from event $e-1$ to event $e$ as $\tau_{e}$ (i.e., $\tau_{e}= t_e-t_{e-1}$) and specify the distribution of candidate timestamps with sender-specfic mean $\mu_{ie}$. Following the generalized linear model (GLM) framework \citep{nelder1972generalized}, we assume the mean and variance of the $\tau_{ie}$ satistify~\looseness=-1
	\begin{equation}
		\begin{aligned}
			E(\tau_{ie}) &= \mu_{ie},\\
			V(\tau_{ie}) &= V(\mu),
		\end{aligned}
	\end{equation}
	where $\tau_{ie}$ here is a positive real number. Possible choices of distribution include exponential, Weibull, gamma, and log-normal distributions, which are commonly used in time-to-event modeling \citep{rao2000applied,rizopoulos2012joint}. Based on the specific distribution, we may need other latent variables to draw the time increment, to account for the variance of time increments, beyond the coefficients for the features used to model the rate. $V(\mu)$---e.g., the shape parameter $k$ for the Weibull, the shape parameter $\theta$ for the gamma, and the variance parameter $\sigma_\tau^2$ for the log-normal. We use $f_\tau(\cdot; \mu, V(\mu))$ and $F_\tau(\cdot; \mu, V(\mu))$ to denote the probability density function (p.d.f) and cumulative density function (c.d.f), respectively, with mean $\mu$ and variance $V(\mu)$.~\looseness=-1

	\subsection{Senders, receivers, and timestamps}\label{subsec:Observed}
\begin{figure}[!t]
			\centering
			\includegraphics[width=0.625\textwidth]{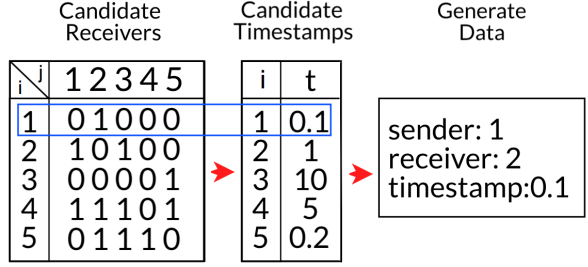}	
			\caption {An illustrative example of the generative process of the HEM.}
			\label{figure:diagram}
		\end{figure}

	Finally, our model assumes that the observed sender, receivers, and timestamp of hyperedge event $e$ are generated by selecting the sender--receiver-set pair with the smallest time increment \citep{snijders1996stochastic}:
	\begin{equation}
		\begin{aligned}
			s_e &= \mbox{argmin}_{i}(\tau_{ie}),\\
			\boldsymbol{r}_e &= \boldsymbol{u}_{s_e e},\\
			t_e &=t_{e-1} + \tau_{s_e e}.
		\end{aligned}
	\end{equation}
	\begin{algorithm}[!t]
		\spacingset{1}
		\SetAlgoLined
		\caption{Generative process: one sender and one or more receivers}
		\begin{algorithmic}
			\STATE \textbf{Input}: number of events and nodes $(E, A)$, covariates $(\boldsymbol{x}, \boldsymbol{y})$, and coefficients $(\boldsymbol{b}, \boldsymbol{\eta})$
			\vskip 0.1in
			\FOR{$e=1$ to $E$}
			\FOR{$i=1$ to $A$}
			\FOR{$j=1$ to $A$ ($j \neq i$)}
			\STATE	set $\lambda_{iej} = {\boldsymbol{b}}^{\top}\boldsymbol{x}_{iej}$
			\ENDFOR
			\STATE	draw $\boldsymbol{u}_{ie}  \sim
			\mbox{MB}_G(\boldsymbol{\lambda}_{ie})$
			\STATE		set $\mu_{ie} = g^{-1}(\boldsymbol{\eta}^\top \boldsymbol{y}_{ie})$
			\STATE		draw $\tau_{ie} \sim f_\tau(\mu_{ie}, V(\mu))$
			\ENDFOR
			
			\IF{$n \geq 2$ tied events} 
			\STATE	set $s_e,\ldots, s_{e+n-1}=\mbox{argmin}_{i}(\tau_{ie}),$
			\STATE	set $\boldsymbol{r}_e=\boldsymbol{u}_{s_e e},\ldots,\boldsymbol{r}_{e+n-1}=\boldsymbol{u}_{s_{e+n-1} e}$
			\STATE	set $t_e, \ldots, t_{e+n-1}=t_{e-1} + \min_i\tau_{ie}$
			\STATE		jump to $e = e+n$
			\ELSE
			\STATE	set $s_e= \mbox{argmin}_{i}(\tau_{ie})$
			\STATE		set $\boldsymbol{r}_e = \boldsymbol{u}_{s_e e}$
			\STATE	set $t_e =t_{e-1} + \min_i\tau_{ie}$
			\ENDIF
			\ENDFOR
		\end{algorithmic}
		\label{alg:generative}
	\end{algorithm}
		Therefore, the HEM assumes a sender-driven process---i.e., the receivers and timestamp of an event are jointly determined by the sender's urgency to direct the event to those selected receivers. Note that our generative process allows for tied events. In the case of tied events---i.e., multiple senders draw exactly the same candidate timestamps---we assume that all events are generated and occur simultaneously. Algorithm \ref{alg:generative} summarizes the entire generative process for hyperedge events with one sender and one or more receivers, and Figure \ref{figure:diagram} presents an illustrative example on how the $e^{\textrm{th}}$ event is generated, assuming $t_{e-1} = 0$ and $A=5$.~\looseness=-1	
		
	\section{Posterior inference}\label{sec:inference}
	In this section we describe how we invert the generative process to obtain the posterior distribution over the latent variables---candidate receivers $\{\boldsymbol{u}_e\}_{e=1}^E$, coefficients for receiver selection features $\boldsymbol{b}$, and coefficients for event timing features $\boldsymbol{\eta}$---conditioned on the observed data $\{(s_e, \boldsymbol{r}_e, t_e)\}_{e=1}^E$, covariates $\{(\boldsymbol{x}_e, \boldsymbol{y}_e)\}_{e=1}^E$, and hyperparamters $(\boldsymbol{\mu}_b, \Sigma_b, \boldsymbol{\mu}_\eta, \Sigma_\eta)$. We draw the samples using Markov chain Monte Carlo (MCMC) methods, repeatedly resampling the value of each latent variable from its conditional posterior via a Metropolis-within-Gibbs sampling algorithm. In the next subsection, we provide each latent variable's conditional posterior along with pseudocode of MCMC in Algorithm \ref{alg:MCMC}. We also evaluate the correctness of both our mathematical derivations and software implemenation using the prior--posterior simulator test of \cite{geweke2004getting}.
	
	\subsection{Conditional posteriors}\label{subsec:conditionaldist}
	\subsubsection{Candidate receivers}
	In our model, direct computation of the posterior densities for the latent variables $\boldsymbol{b}$ and $\boldsymbol{\eta}$---i.e., $\Pr(\boldsymbol{b}|\,\boldsymbol{x},\boldsymbol{s}, \boldsymbol{r},\boldsymbol{t})$ and $\Pr(\boldsymbol{\eta}|\,\boldsymbol{y},\boldsymbol{s}, \boldsymbol{r},\boldsymbol{t})$---is not possible. However, it is possible to augment the data by candidate receivers $\boldsymbol{u}$ such that we can obtain their conditional posterior by conditioning on samples of $\boldsymbol{u}$. 
	This approach---i.e., ``data augmentation"---is commonly used throughout Bayesian statistics \citep{tanner1987calculation,neal2015exact}. Since $u_{iej}$ is a binary random variable, it may be sampled from a Bernoulli distribution with probability $p_{iej} =\frac{\exp(\lambda_{iej})}{\exp(\lambda_{iej})+\text{I}(\lVert\boldsymbol{u}_{ie\backslash j}\rVert_1 > 0 )}$, since
	\begin{equation}
		\begin{aligned}
			&\Pr(u_{iej}=1| \,\boldsymbol{u}_{ie\backslash j}, \boldsymbol{b}, \boldsymbol{x},\boldsymbol{s}, \boldsymbol{r},\boldsymbol{t}) \propto \exp(\lambda_{iej}) \\
			&\Pr(u_{iej}=0|\, \boldsymbol{u}_{ie\backslash j},\boldsymbol{b}, \boldsymbol{x},\boldsymbol{s}, \boldsymbol{r},\boldsymbol{t})\propto \text{I}(\lVert\boldsymbol{u}_{ie\backslash j}\rVert_1 > 0 ),
		\end{aligned}
		\label{eqn:latentreceiver}
	\end{equation}
	where the subscript ``$\backslash j$'' denotes a quantity excluding data from position $j$ and $\text{I}(\cdot)$ is the indicator function that prevents empty receiver sets. 
	\subsubsection{Coefficients for receiver selection features}
	Unlike the candidate receivers above, the conditional posterior for $\boldsymbol{b}$ does not have a closed form; however $\boldsymbol{b}$ may instead be re-sampled using the Metropolis--Hastings (MH) algorithm. Assuming an uninformative prior (i.e., $N({0},\infty)$), the conditional posterior for $\boldsymbol{b}$ is proportional to~\looseness=-1
	\begin{equation}
		\Pr(\boldsymbol{b}| \,\boldsymbol{u}, \boldsymbol{x}, \boldsymbol{s}, \boldsymbol{r},\boldsymbol{t})\propto \prod_{e=1}^E
		\prod_{i=1}^A \frac{1}{Z(\boldsymbol{\lambda}_{ie})}\exp\Big(\mbox{log}\big(\text{I}( \lVert \boldsymbol{u}_{ie}\rVert_1 > 0)\big) + \sum\limits_{j \neq i} \lambda_{iej}u_{iej}\Big).
		\label{eqn:latentedge}
	\end{equation}
	\subsubsection{Coefficients for event timing features}
	Likewise, we use the MH algorithm to update the latent variable $\boldsymbol{\eta}$. Assuming an uninformative prior $\boldsymbol{\eta}$ (i.e., $N({0},\infty)$), the conditional posterior for an untied event case is proportional to~\looseness=-1
	\begin{equation}
		\Pr(\boldsymbol{\eta}|\, \boldsymbol{u}, \boldsymbol{y},\boldsymbol{s}, \boldsymbol{r},\boldsymbol{t})\propto \prod_{e=1}^E\Big(f_{\tau}(\tau_{e}; \mu_{s_e e}, V(\mu))\times \prod_{i\neq s_e}\big(1-F_{\tau}(\tau_{e}; \mu_{ie}, V(\mu)) \big)\Big),
		\label{eqn:latenttime}
	\end{equation}
	where $f_{\tau}(\tau_{e}; \mu_{s_e e}, V(\mu))$ is the probability that the $e^{\textrm{th}}$ observed time increment comes from the specified distribution $f_\tau(\cdot)$ with the observed sender's mean $\mu_{s_e e}$, and $\prod_{i\neq s_e}\big(1-F_{\tau}(\tau_{e}; \mu_{ie},V(\mu)) \big)$ is the probability that the rest of (unobserved) senders for event $e$ all draw time increments greater than $\tau_e$. Moreover, under the existence of tied events, the conditional posterior of $\boldsymbol{\eta}$ is written as proportional to
	\begin{equation}
		\begin{aligned}
			\Pr(\boldsymbol{\eta}|\, \boldsymbol{u}, \boldsymbol{y},\boldsymbol{s}, \boldsymbol{r},\boldsymbol{t})&\propto \prod_{m=1}^M\Big(\prod_{e:t_e=t_m^*}f_{\tau}(t_m^*-t_{m-1}^*; \mu_{s_e e}, V(\mu)) \\&\times \prod_{i \notin \{s_e\}_{e:t_e=t_m^*}}\big(1-F_{\tau}(t_m^*-t_{m-1}^*; \mu_{ie}, V(\mu)) \big)\Big),
		\end{aligned}
		\label{eqn:latenttime2}
	\end{equation}
	where $t_1^*,\ldots,t_M^*$ are the unique timepoints across $E$ events ($M \leq E$). If $M=E$ (i.e., no tied events), equation (\ref{eqn:latenttime2}) reduces to equation (\ref{eqn:latenttime}). Note that when we have the latent variable to quantify the variance in time increments $V(\mu)$ (based on the choice of timestamp distribution in Section \ref{subsec:Time}), we also use equation (\ref{eqn:latenttime}) (or equation (\ref{eqn:latenttime2}) in case there exist tied events) for the additional MH update---e.g., $\Pr(k|\, \boldsymbol{\eta},\boldsymbol{u}, \boldsymbol{y},\boldsymbol{s}, \boldsymbol{r},\boldsymbol{t})$ for Weibull, $\Pr(\theta|\, \boldsymbol{\eta},\boldsymbol{u}, \boldsymbol{y},\boldsymbol{s}, \boldsymbol{r},\boldsymbol{t})$ for gamma, and  $\Pr(\sigma^2_\tau| \,\boldsymbol{\eta},\boldsymbol{u}, \boldsymbol{y},\boldsymbol{s}, \boldsymbol{r},\boldsymbol{t})$ for log-normal.~\looseness=-1
		\begin{algorithm}[!t]
			\spacingset{1}
			\SetAlgoLined
			\caption{MCMC algorithm}
			\begin{algorithmic}
				\STATE \textbf{Input}: number of outer and inner iterations $(O, I_1, I_2)$ and initial values of $(\boldsymbol{u}, \boldsymbol{b}, \boldsymbol{\eta})$
				\vskip 0.1in
				\FOR{$o=1$ to $O$}
				\FOR{$e=1$ to $E$}
				\FOR{$i = 1$ to $A$}
				\FOR{$j = 1$ to $A$ ($j \neq i$)}
				\STATE update $u_{iej}$ using Gibbs update ---equation (\ref{eqn:latentreceiver})
				\ENDFOR
				\ENDFOR
				\ENDFOR
				\FOR{$n=1$ to $I_1$}
				\STATE update $\boldsymbol{b}$ using MH algorithm---equation (\ref{eqn:latentedge})
				\ENDFOR
				\FOR{$n=1$ to $I_2$}
				\STATE update $\boldsymbol{\eta}$ using MH algorithm---equation (\ref{eqn:latenttime}) or (\ref{eqn:latenttime2}) 
				\ENDFOR
				\IF {extra parameter for $V(\mu)$} 
				\STATE update the variance parameter using MH algorithm---equation (\ref{eqn:latenttime}) or (\ref{eqn:latenttime2}) 
				\ENDIF
				\ENDFOR
				\STATE	summarize the results with the last chain of $\boldsymbol{b}$ and $\boldsymbol{\eta}$
			\end{algorithmic}
			\label{alg:MCMC}
		\end{algorithm}

	\subsection{Getting it Right (GiR) test} \label{subec:GiR}
	Software development is integral to the objective of applying our model to real world data. Code review is a valuable process in any research computing context, and the prevalence of software bugs in statistical software is well documented \citep[e.g., ][]{altman2004numerical,mccullough2009accuracy}.  With highly complex models such as the HEM, there are many ways in which software bugs can be introduced and go unnoticed. As such, we present a joint analysis of the integrity of our generative model, sampling equations, and software implementation. 
	
	\cite{geweke2004getting} introduced the ``Getting it Right'' (GiR) test---a joint distribution test of posterior simulators which can detect errors in sampling equations as well as software bugs---and it has been used to test the implementation of Bayesian inference algorithms \citep{zhao2016bayesian}.  The test involves comparing the distributions of variables simulated from two joint distribution samplers, which we call ``forward" and ``backward" samplers. The ``forward" sampler draws joint samples of the latent and observable variables from the prior. The ``backward" sampler begins by first drawing a joint sample of the latent and observed variables from the prior. It then alternates between re-sampling the latent variables, conditioned on the observable variables, from the MCMC transition operator, and then re-sampling the observable variable, conditioned on the latent variables, from the model likelihood. If the MCMC transition operator is correctly derived and implemented, this process should asymptotically generate joint samples of the latent and observable variables from the prior, like the forward sampler.
		
	In the forward sampler, both observable and unobservable variables are generated using Algorithm \ref{alg:generative}. In the backward samples, unobservable variables are generated using the sampling equations for inference, which we derived in Section \ref{subsec:conditionaldist}. For each forward and backward sample that consists of $E$ number of events, we save these statistics:\\
	\begin{itemize}
		\item[1.] Mean of observed receiver sizes $ \lVert \boldsymbol{r}_{e} \rVert_1 $ across $e=1,\ldots,E$,
		\item[2.] Variance of observed receiver sizes $ \lVert \boldsymbol{r}_{e} \rVert_1 $ across $e=1,\ldots,E$,
		\item[3.] Mean of time increments $\tau_e$ across $e=1,...,E$,
		\item[4.] Variance of time increments $\tau_e$ across $e=1,...,E$,
		\item[5.] $b_p$ value used to generate the samples $p = 1,...,P$,
		\item[6.] $\eta_q$ value used to generate the samples $q = 1,...,Q$,
		\item[7.] $\sigma^2_\tau$ value used to generate the samples in log-normal distribution
	\end{itemize}

\begin{figure}[!t]
			\centering
			\begin{subfigure}[b]{0.2425\textwidth}
				\caption{Mean of $\lVert \boldsymbol{r}_{e} \rVert_1 $}
				\includegraphics[width=\textwidth]{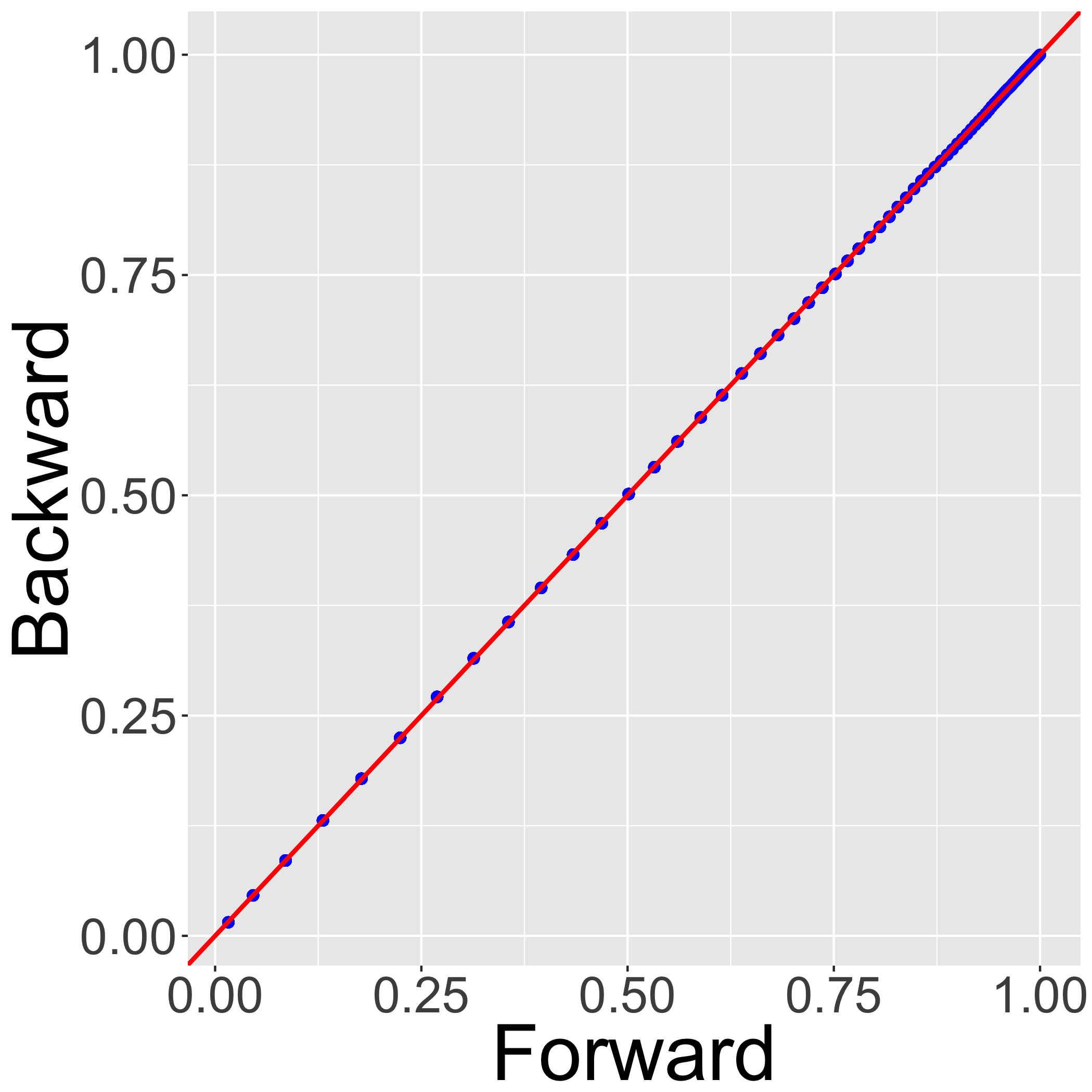}	
			\end{subfigure}
			\begin{subfigure}[b]{0.2425\textwidth}
				\caption{Variance of $\lVert \boldsymbol{r}_{e} \rVert_1 $}
				\includegraphics[width=\textwidth]{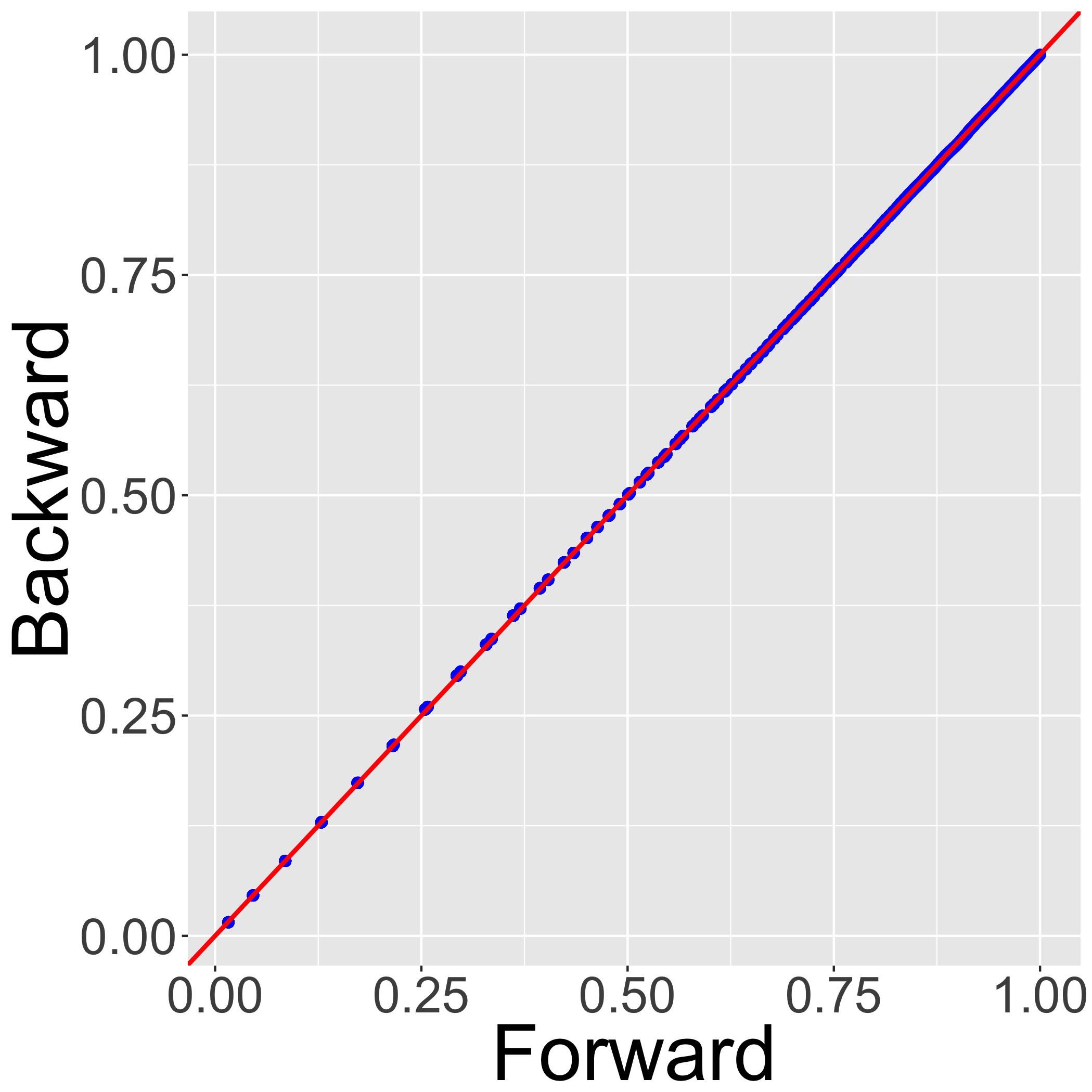}	
			\end{subfigure}
			\begin{subfigure}[b]{0.2425\textwidth}
				\caption{Mean of $\tau_e$}
				\includegraphics[width=\textwidth]{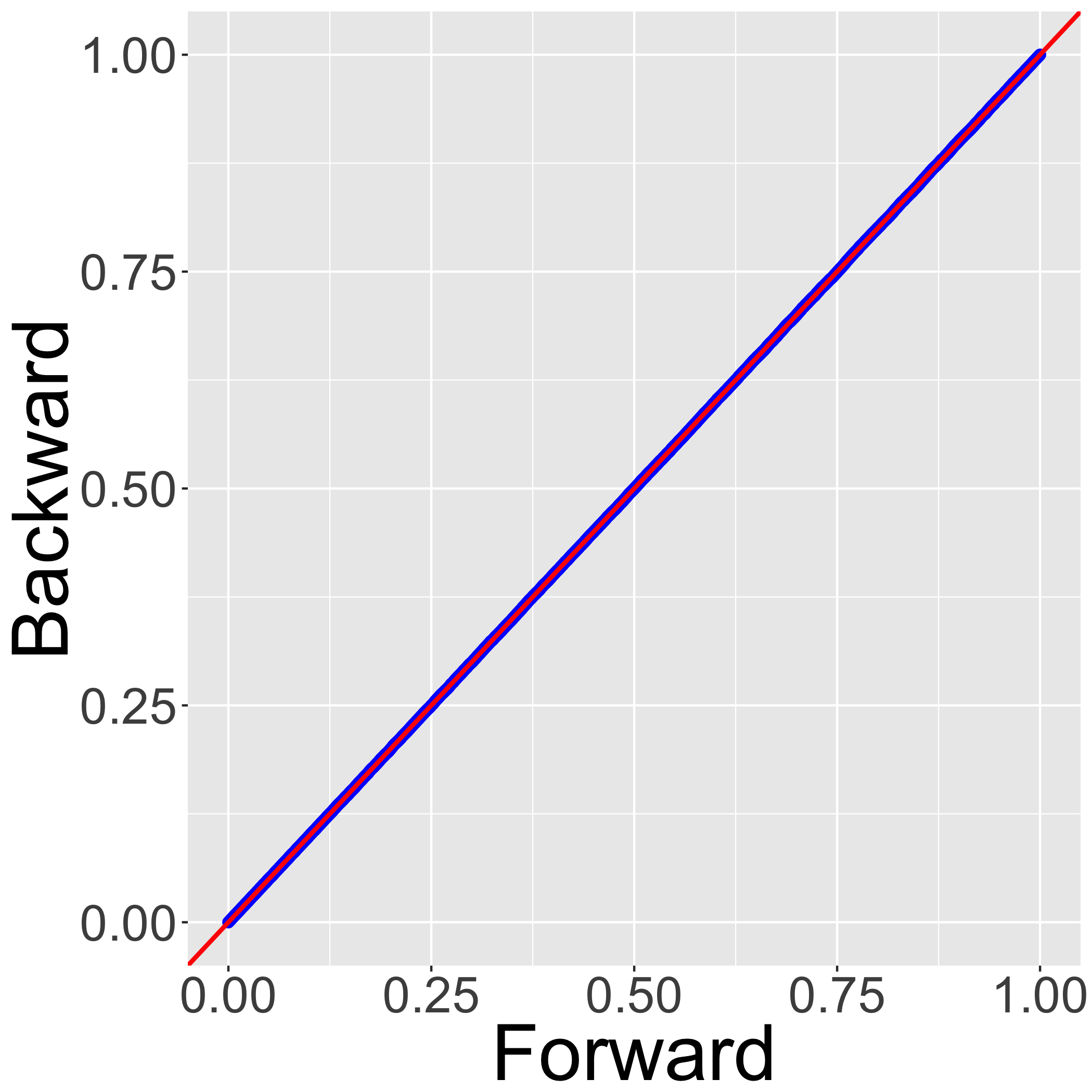}	
			\end{subfigure}
			\begin{subfigure}[b]{0.2425\textwidth}
				\caption{Variance of $\tau_e$}
				\includegraphics[width=\textwidth]{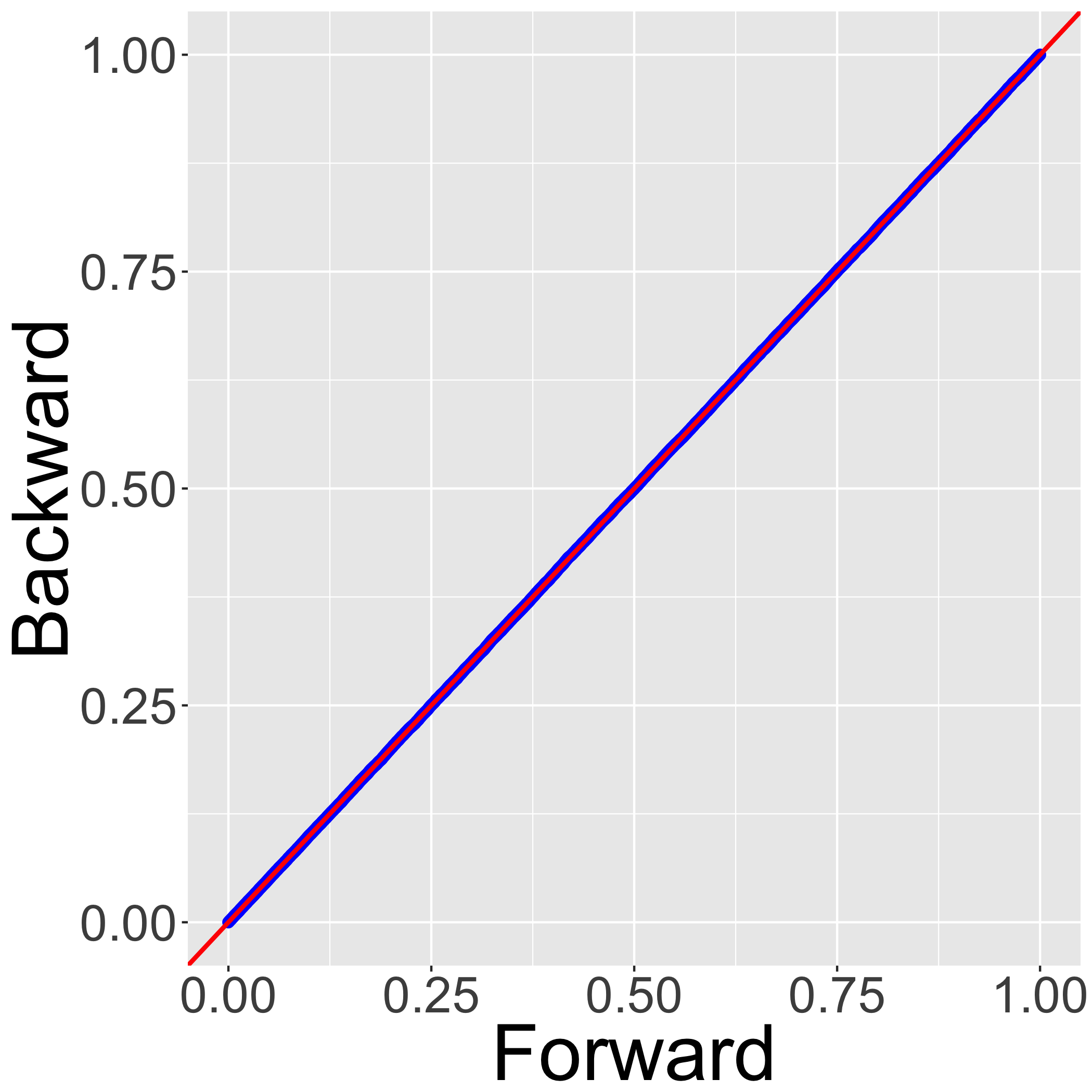}	
			\end{subfigure}
			\begin{subfigure}[b]{0.2425\textwidth}
				\caption{Value of $b_1$}
				\includegraphics[width=\textwidth]{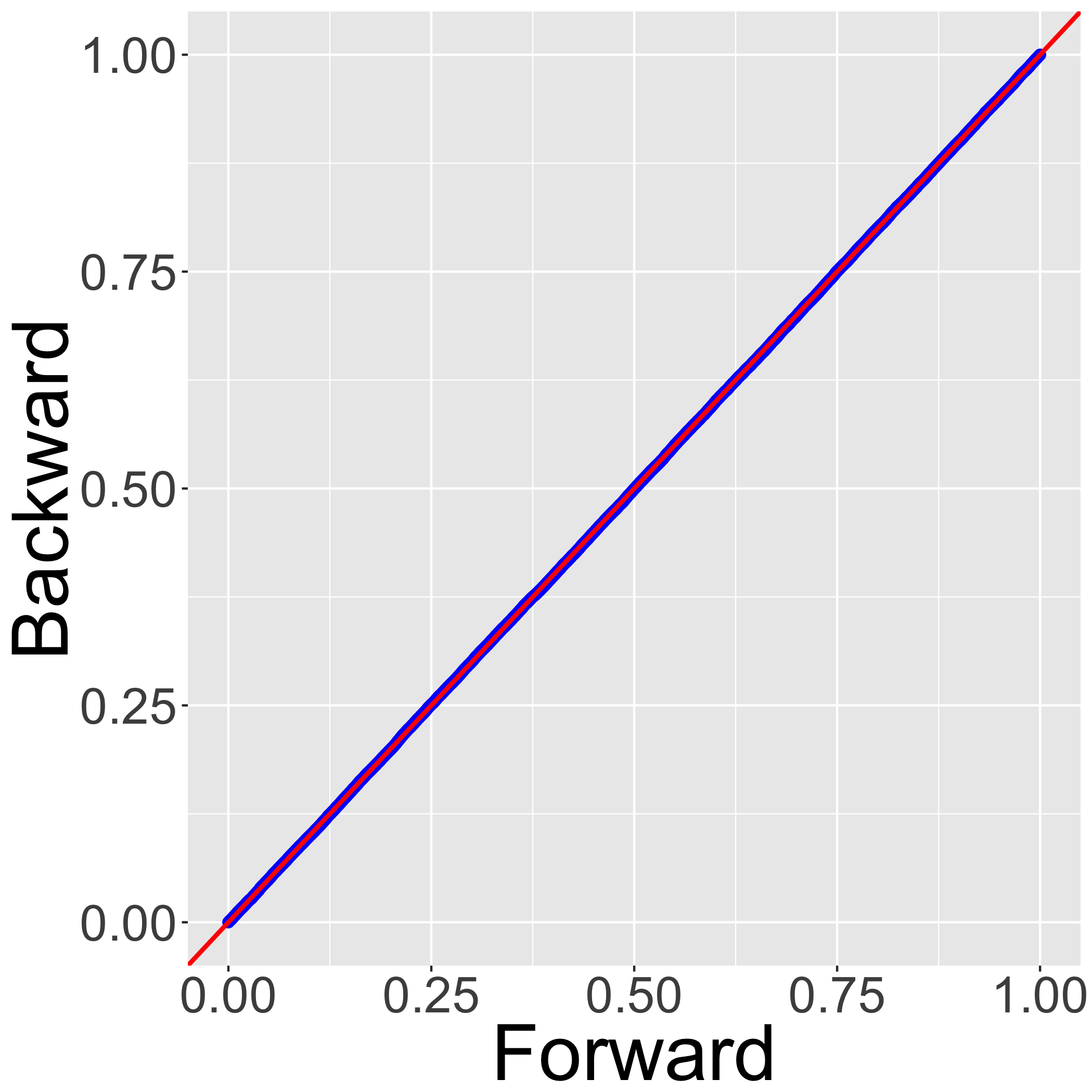}	
			\end{subfigure}
			\begin{subfigure}[b]{0.2425\textwidth}
				\caption{Value of $b_2$}
				\includegraphics[width=\textwidth]{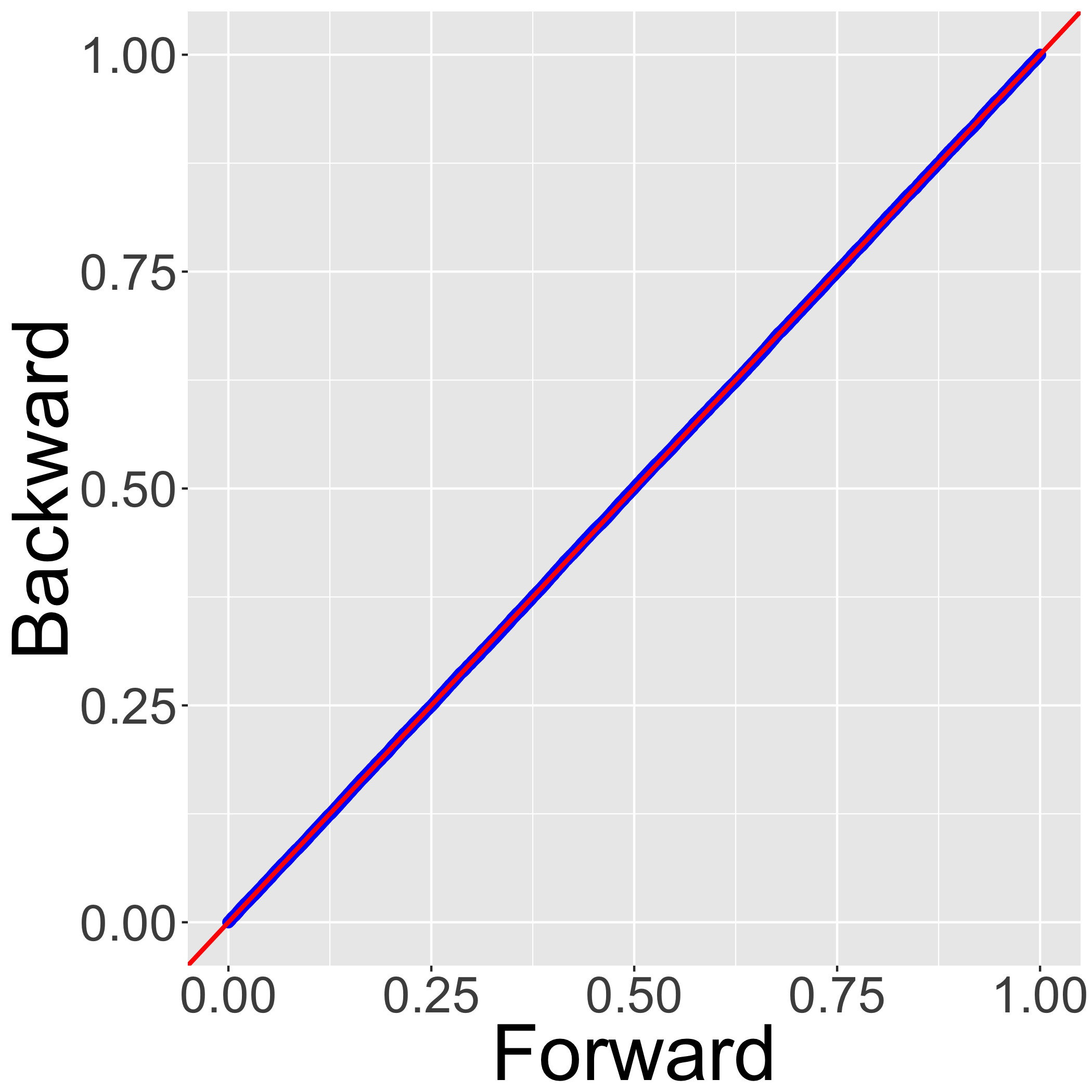}	
			\end{subfigure}   	   	   	   	   	   	   	
			\begin{subfigure}[b]{0.2425\textwidth}
				\caption{Value of $b_3$}
				\includegraphics[width=\textwidth]{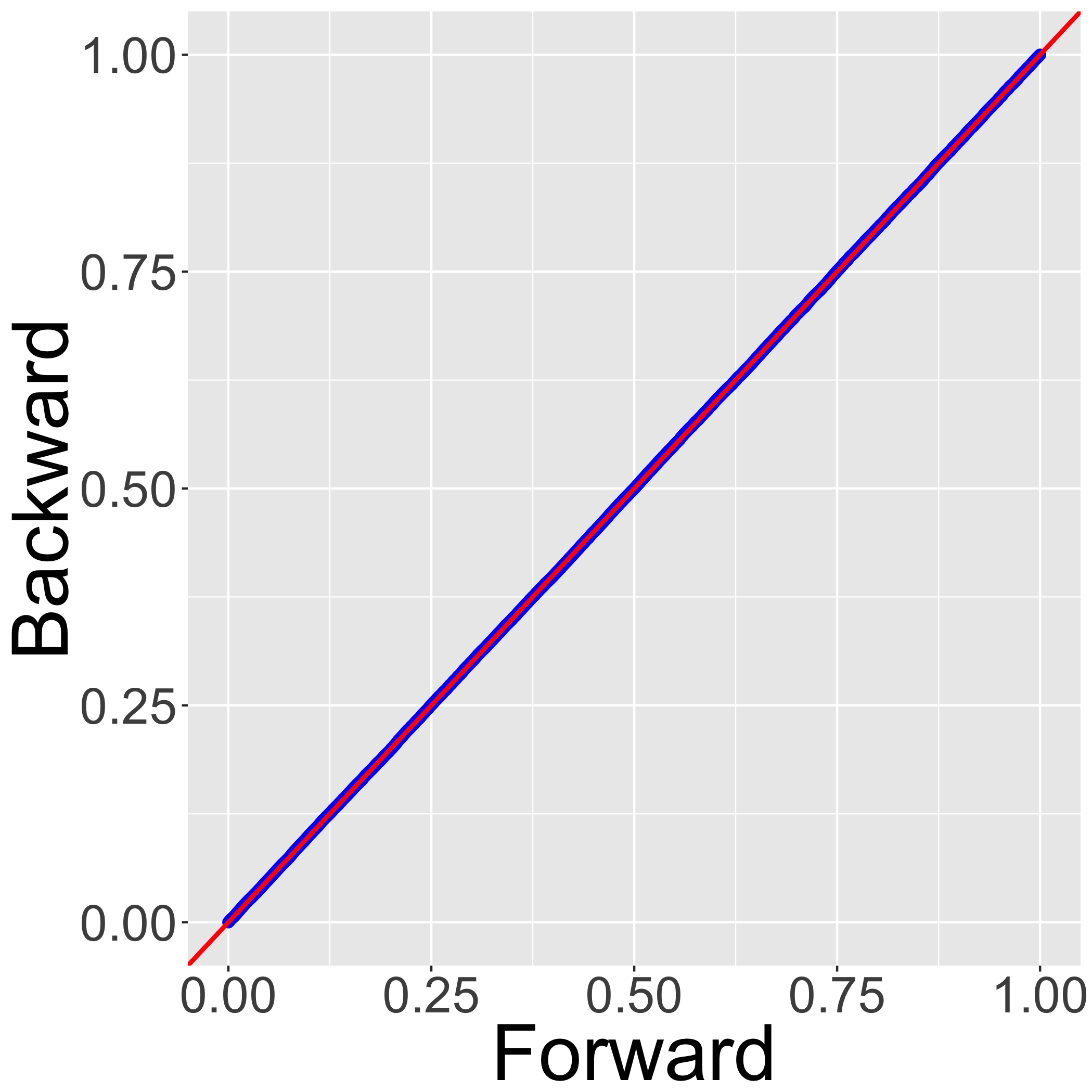}	
			\end{subfigure}   	   		   	   	
			\begin{subfigure}[b]{0.2425\textwidth}
				\caption{Value of $b_4$}
				\includegraphics[width=\textwidth]{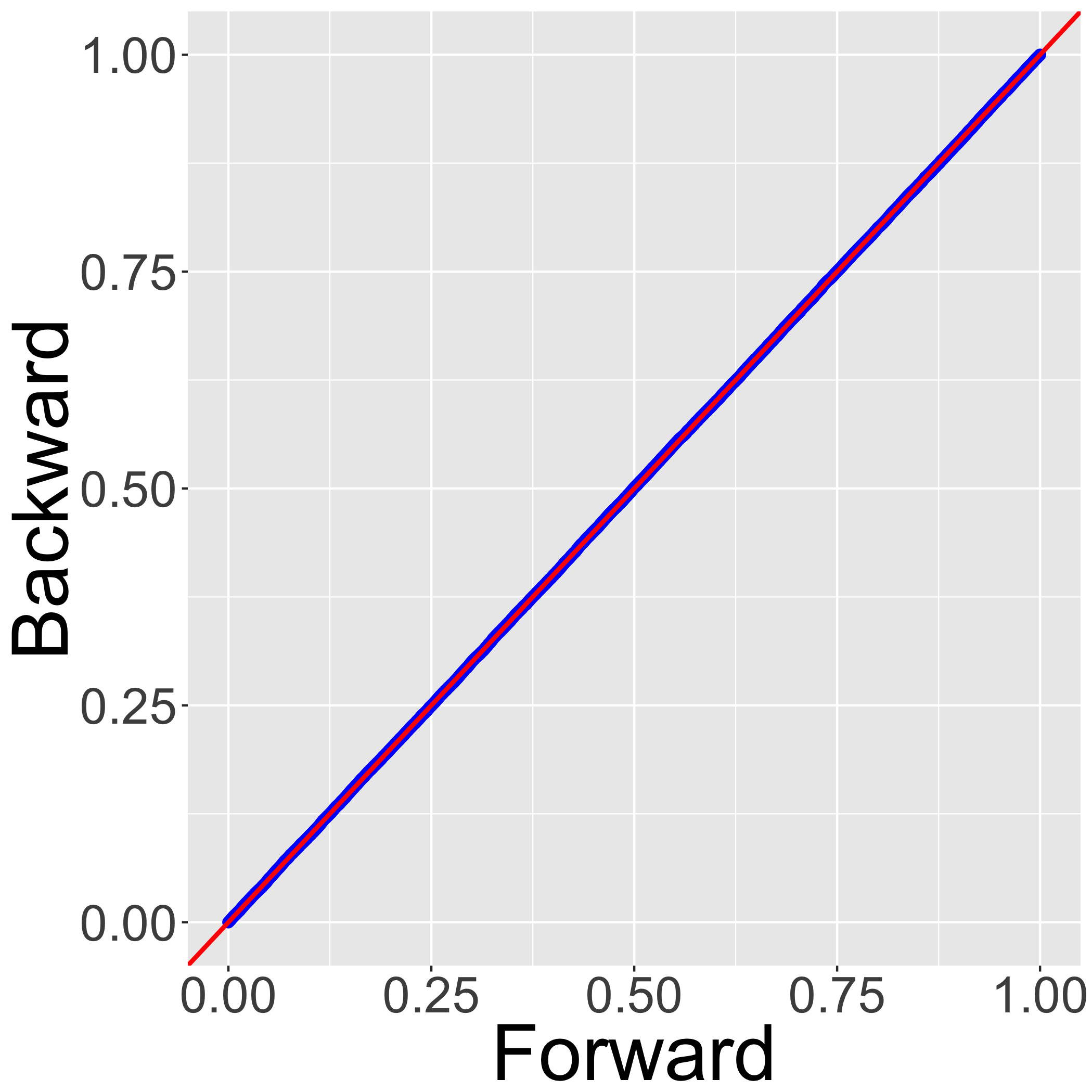}	
			\end{subfigure}   	   		   	   	
			\begin{subfigure}[b]{0.2425\textwidth}
				\caption{Value of $\eta_1$}
				\includegraphics[width=\textwidth]{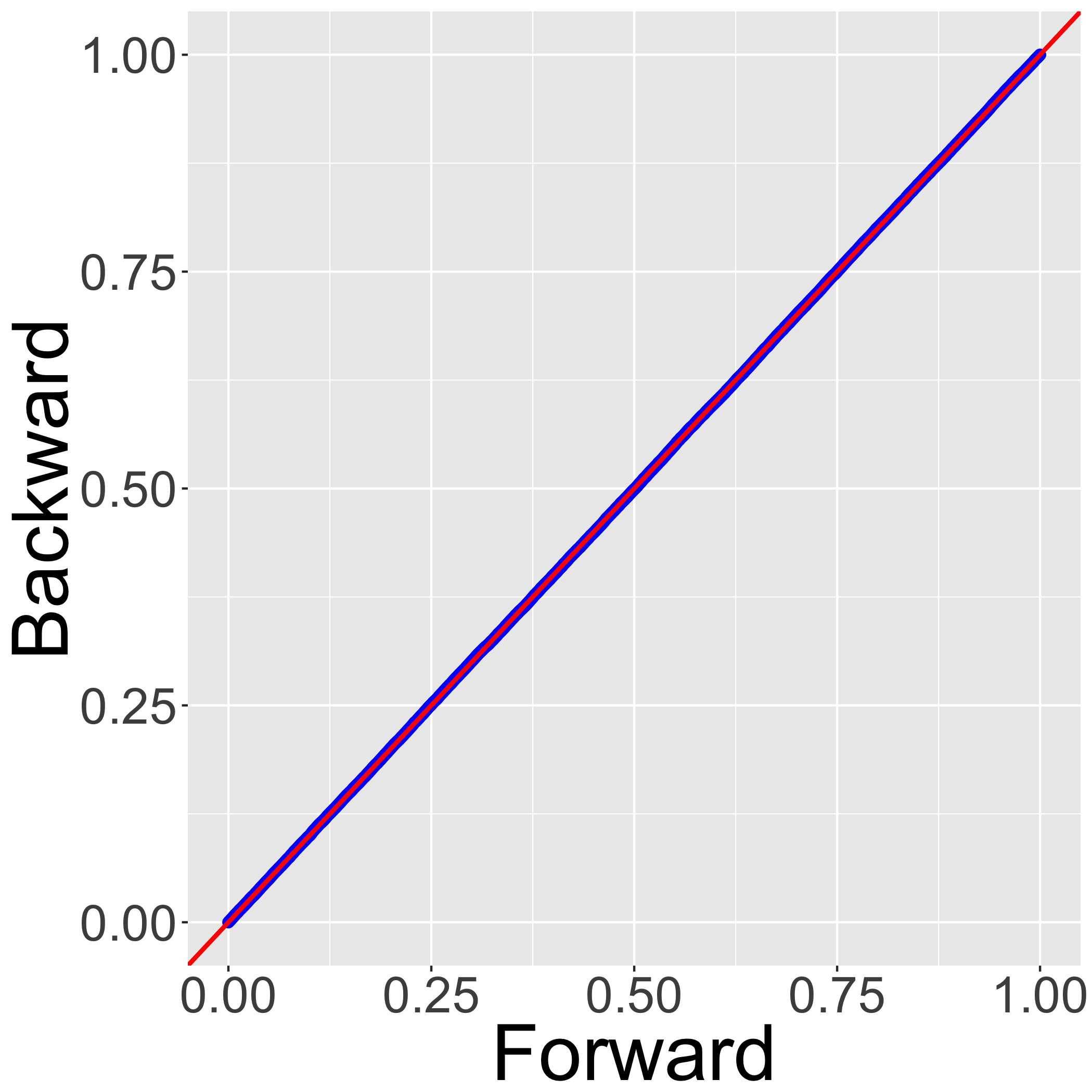}	
			\end{subfigure}   	   		   	   	
			\begin{subfigure}[b]{0.2425\textwidth}
				\caption{Value of $\eta_2$}
				\includegraphics[width=\textwidth]{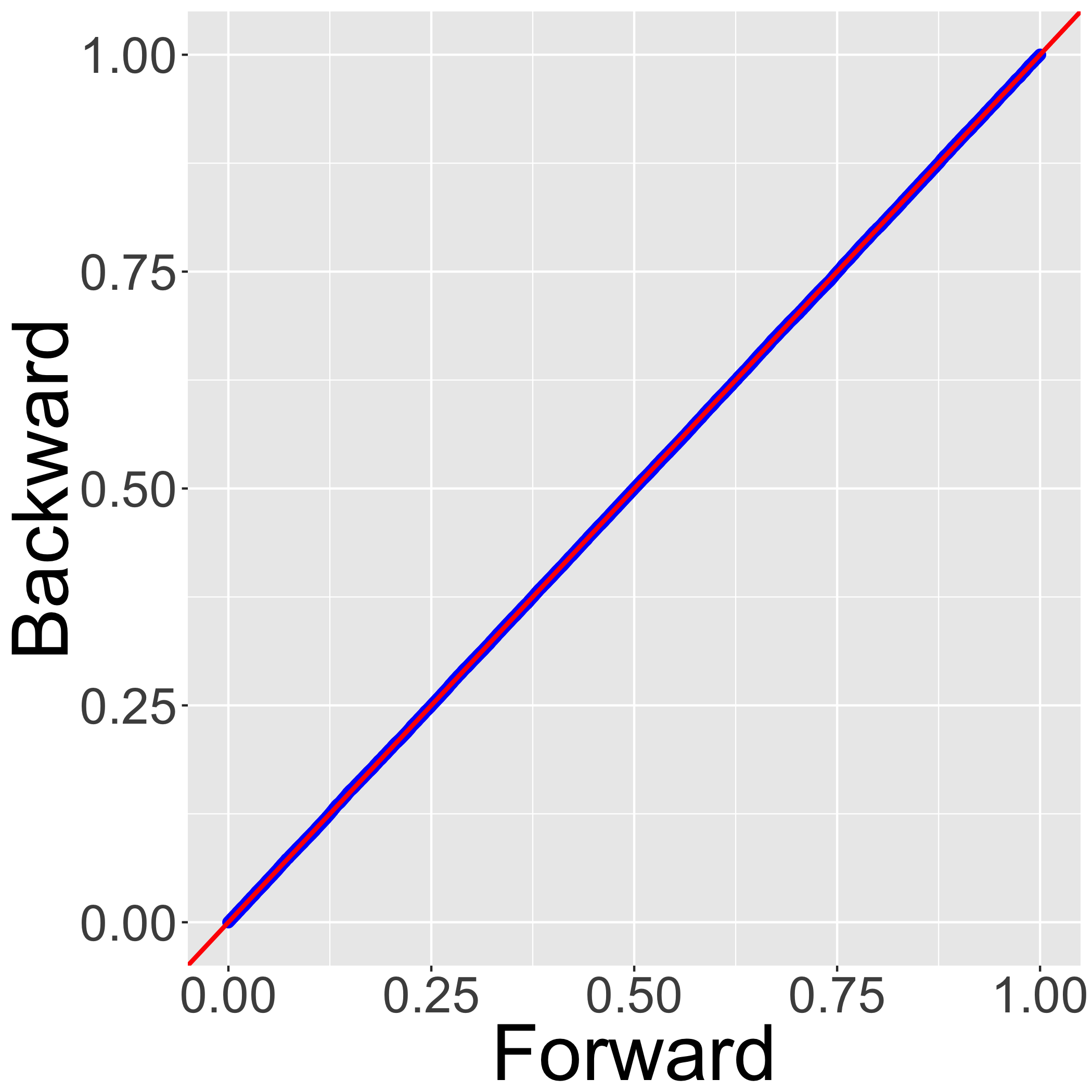}	
			\end{subfigure}   	   		   	   	
			\begin{subfigure}[b]{0.2425\textwidth}
				\caption{Value of $\eta_3$}
				\includegraphics[width=\textwidth]{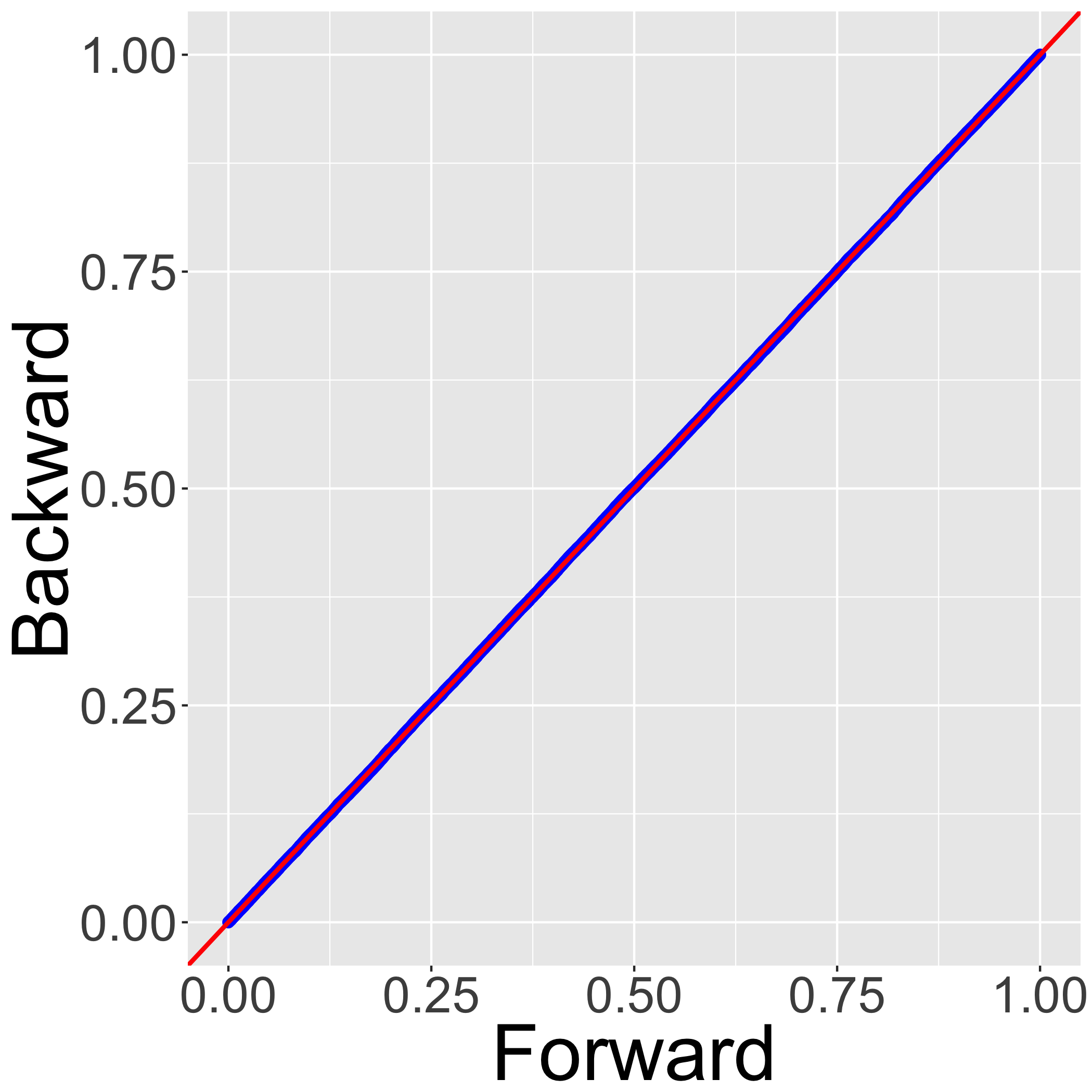}	
			\end{subfigure}   	   		   	   	  		   	   	 	
			\begin{subfigure}[b]{0.2425\textwidth}
				\caption{Value of $\sigma^2_\tau$}
				\includegraphics[width=\textwidth]{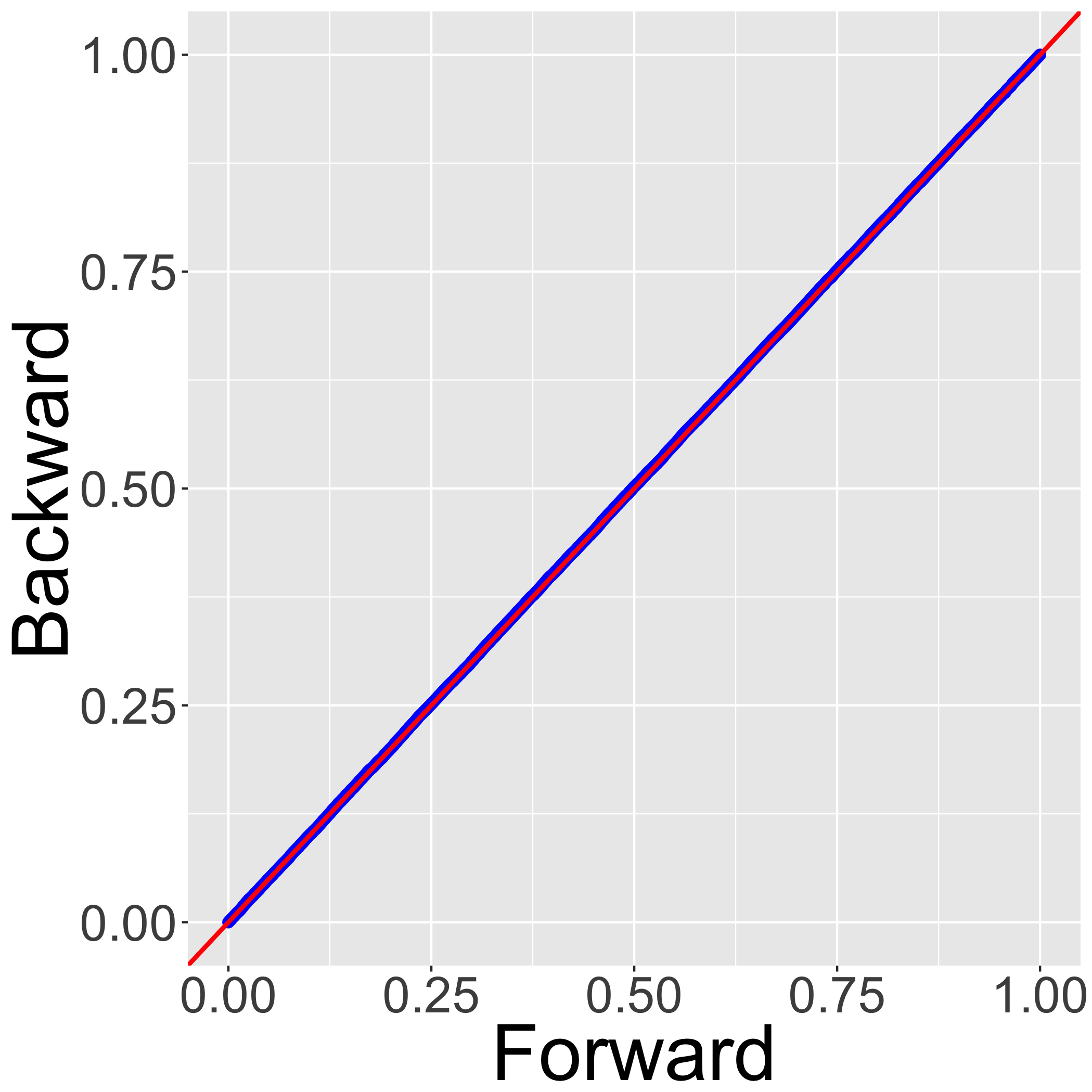}	
			\end{subfigure}   	   		   	   	   	   		   	   	   		   	   	
			\caption{Probability--probability (P--P) plots for the GiR test statistics.}
			\label{figure:GiRplot}
		\end{figure}

	To keep the computational burden of re-running thousands of rounds of inference manageable, we run the GiR using a relatively small artificial sample, consisting of $E=100$ events, $A=5$ nodes, $P=4$ number of receiver selection features, and $Q=3$ number of event timing features per each forward or backward sampler, using log-normal distibution for the time increments $f_\tau$. We generated $10^5$ sets of forward and backward samples, and then calculated 1,000 quantiles for each of the statistics. We also calculated t-test and Mann-Whitney test p-values in order to test for differences in the distributions generated in the forward and backward samples. Before we calculated these statistics, we thinned our samples by taking every 9th sample starting at the 10,000th sample for a resulting sample size of 10,000, in order to reduce the autocorrelation in the Markov chains. In each case, if we observe a large p-value, this gives us evidence that the distributions generated under forward and backward sampling have the same locations. We depict the GiR results using probability--probability (P--P) plots, in which the empirical CDF values of the forward and backward samples are plotted on the $x$ and $y$ axes, respectively. If the two samples are from equivalent distributions, the empirical CDF values should line up on a line with zero $y$-intercept, and unit slope (i.e., a 45-degree line). The GiR test results are depicted in Figure \ref{figure:GiRplot}. These results indicate that our sampling equations and software implementation pass the test on every statistic.
	\section{Application to email data}\label{sec:Emails}
	We now present a case study applying our method to Montgomery county government email data.
	Our data come from the North Carolina county government email dataset collected by \cite{ben2017transparency} that includes internal email corpora covering the inboxes and outboxes of managerial-level employees of North Carolina county governments. Out of over twenty counties, we chose Montgomery County to (1) test our model using data with a large proportion of hyperedges (16.76\%), all of which are emails sent from one sender to two or more receivers, and (2) limit the scope of this initial application. The Montgomery County email network contains 680 emails, sent and received by 18 department managers over a period of 3 months (March--May) in 2012. For this case study,
	we formulate our model specification through definitions of the receiver selection features $\boldsymbol{x}$ and event timing features $\boldsymbol{y}$. We then report a suite of experiments---out-of-sample prediction for model selection and posterior predictive checks---that illustrate how alternative formulations of the HEM can be compared, and evaluate how well our model recovers the distribution of the observed data. Finally, we demonstrate an exploratory analysis of Montgomery County email data using the model estimates to discover substantively meaningful patterns in organizational communication networks.
	\subsection{Covariates}\label{subsec:Covariates_email}
	\subsubsection{Receiver selection features}
\begin{figure}[!t]
		\centering
		\includegraphics[width=0.75\textwidth]{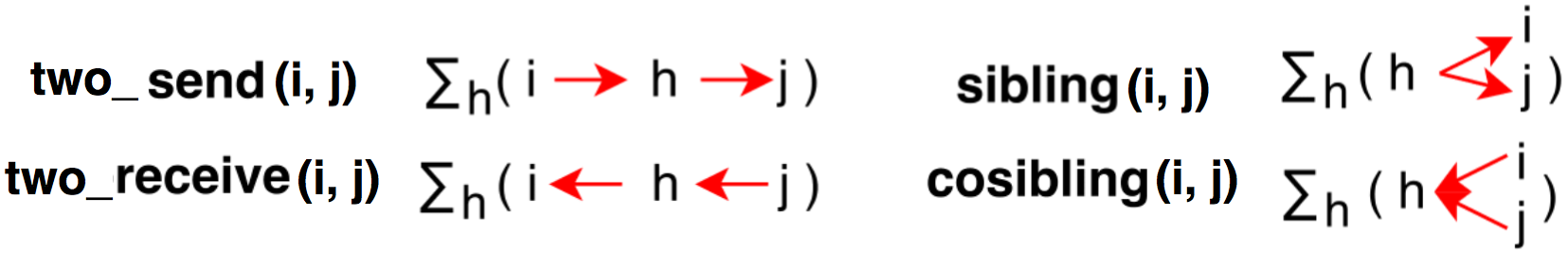}	
		\caption {Visualization of triadic statistics: two\_\,send, two\_\,receive, sibling, and cosibling.}
		\label{figure:netstats}
	\end{figure}

	A primary purpose of any network model is to use the posterior distributions to learn which features predict and/or explain edge formation (e.g., is edge formation reciprocal, are edges more likely to be formed among nodes with the same gender). This email application specifically gives rise to the following question: ``To what extent are nodal, dyadic or triadic network effects relevant to predicting future emails?" As an illustrative example, we form the receiver selection features $\boldsymbol{x}$ for Montgomery County email data using nodal, dyadic, and triadic covariates. First, as we want to test whether gender plays a role in receiver selection process, we include three nodal covariates---the gender information of sender and receiver, and their homophily indicator (i.e., an indicator of whether the sender and receiver are of the same gender).  Additionally, we include four interval-based nodal network covariates---outdegree of sender (i.e., the number of emails sent), indegree of receiver (i.e., the number of emails received), hyperedge size of sender (i.e., the number of total receivers directed from the sender), and the interaction between (i.e., scalar product of) outdegree and hyperedge size---to study the effect of nodal behaviors on future interactions. For dyadic and triadic network effects, we employ the network statistics in \cite{PerryWolfe2012} and summarize past interaction behaviors based on the time interval prior to and including $t_{e-1}$. Specifically, our time interval tracks 7 days prior to the last email was sent $l_e= (t_{e-1}-7\mbox{ days}, t_{e-1}]$. For $i \in [A], j\in [A]$, and $e \in [E]$, we define 14 covariates for $\boldsymbol{x}_{iej}$:
	\begin{itemize}
		\item[1.] intercept: ${x}_{iej1} =1$;
		\item[2.] $\mbox{gender}\_\,\mbox{sender}(i)$: ${x}_{iej2} = I(\mbox{gender of sender } i= \mbox{female});$
		\item[3.] $\mbox{gender}\_\,\mbox{receiver}(j)$: ${x}_{iej3} = I(\mbox{gender of receiver } j= \mbox{female});$
		\item[4.] $\mbox{gender}\_\,\mbox{homophily}(i, j)$: ${x}_{iej4} = I({x}_{iej2}={x}_{iej3});$ 	   		
		\item[5.] $\mbox{outdegree}(i)$: ${x}_{iej5} =\sum_{e^\prime: t_{e^\prime} \in l_e} I(s_{e^\prime} = i)$;
		\item[6.] $\mbox{indegree}(r)$: ${x}_{iej6}=\sum_{d^\prime: t_{e^\prime} \in l_e} I(u_{e^\prime j} = 1)$;
		\item[7.] $\mbox{hyperedge}\_\,\mbox{size}(i)$: ${x}_{iej7}=\sum_{e^\prime: t_{e^\prime} \in l_e} \sum_{j=1}^A I(s_{e^\prime} = i)\,I(u_{e^\prime j} = 1)$;
		\item[8.] $\mbox{interaction}(i)$: ${x}_{iej8} = {x}_{iej5}\times{x}_{iej7};$
		\item[9.] $\mbox{send}(i, j)$: ${x}_{iej9}=\sum_{e^\prime: t_{e^\prime} \in l_e} I(s_{e^\prime} = i)\,I(u_{e^\prime j} = 1)$;
		\item[10.] $\mbox{receive}(i,j)$: ${x}_{iej10}=\mbox{send}(j,i)$;
		\item[11.] $\mbox{two}\_\,\mbox{send}(i,j)$: ${x}_{iej11} = \sum_{h \neq i,j} \mbox{send}(i,h)\,\mbox{send}(h,j)$;
		\item[12.] $\mbox{two}\_\,\mbox{receive}(i,j)$: ${x}_{iej12}= \sum_{h \neq i, j} \mbox{send}(h,i)\,\mbox{send}(j,h)$;
		\item[13.] $\mbox{sibling}(i, j)$: ${x}_{iej13}=\sum_{h \neq i, j} \mbox{send}(h,i)\,\mbox{send}(h,j)$;
		\item[14.] $\mbox{cosibling}(i, j)$: ${x}_{iej14}=\sum_{h \neq i,j} \mbox{send}(i,h)\,\mbox{send}(j,h)$;
	\end{itemize}
	where $I(\cdot)$ is an indicator function. The network statistics (5--14) are designed so that their coefficients have a straightforward interpretation. The function ``outdegree$(i)$" and ``indegree$(j)$" measure the gregariousness and popularity effects of the node by counting the number of emails sent from $i$ and received by $j$, respectively, within the last 7 days. The gregariousness effect refers to the tendency for nodes that created many events in the past to continue to do so in the future. The popularity effect refers to the tendency for nodes that were selected as receivers of many events in the past to continue to do so in the future. Moreover, in order to capture the individual tendency of senders to select two or more receivers, we include the statistic ``$\mbox{hyperedge}\_\,\mbox{size}(i)$"---the number of events directed from sender $i$ within last 7 days where events with $n$ number of receivers are counted as $n$ separate events---as a variant of outdegree statistic, accounting for hyperedges. We also include the interaction term, ``$\mbox{interaction}(i)$", between outdegree and hyperedge size. This interaction allows us to model a possible tradeoff between the hyperedge size and the total number of events created by $i$. Dyadic statistics ``send$(i, j)$" and ``receive$(i, j)$" are defined as above such that these covariates measure the number of events directed from $i$ to $j$ and $j$ to $i$, respectively, within the last 7 days. In the example of triadic statistics, the covariate ``$\mbox{two}\_\,\mbox{send}(i, j)$" counts the events involving some node $h$ distinct from $i$ and $j$ such that events from $i$ to $h$ and $h$ to $j$ are both observed within the last 7 days. This statistic captures the tendency for events to close transitive triads (i.e., triads in which $i$ directs to $j$ and $h$, and $j$ directs to $h$). We include other triadic covariates that behave similarly and exhibit analogous interpretations, which are illustrated in Figure \ref{figure:netstats}.
	
	\subsubsection{Event timing features}
For the event timing features $\boldsymbol{y}$, introduced in Section \ref{subsec:Time}, we identify a set of covariates which may affect the time until the next event. Similar to the receiver selection features, we include nodal statistics which are time-invariant (such as gender or manager status) or time-dependent (such as the network statistics used for $\boldsymbol{x}$). In addition, we select some event-specific covariates based on the temporal aspect of the $(e-1)^{\textrm{th}}$ event---e.g., whether the previous email was sent (1) during the weekend and (2) before or past midday (AM/PM)---since we expect the email interactions within county government to be less active during the weekend and in the evening. To be specific, the timestamp statistics are defined as
	\begin{itemize}
		\item[1.] intercept: ${y}_{ie1} =1$;
		\item[2.] $\mbox{gender}(i)$: ${y}_{ie2}=I(\mbox{gender of sender }i= \mbox{female})$;
		\item[3.] $\mbox{manager}(i)$: ${y}_{ie3}=I(\mbox{sender }i \mbox{ is the County Manager})$;
		\item[4.] $\mbox{outdegree}(i)$: ${y}_{ie4} =\sum_{e^\prime: t_{e^\prime} \in l_e} I(s_{e^\prime} = i)$;
		\item[5.] $\mbox{indegree}(i)$: ${y}_{ie5}=\sum_{e^\prime: t_{e^\prime} \in l_e} I(u_{e^\prime i} = 1)$;
		\item[6.] $\mbox{weekend}(e)$: ${y}_{ie6} = I(t_{e-1} \mbox{ is during the } \mbox{weekend})$;
		\item[7.] $\mbox{PM}(e)$: ${y}_{ie7}= I(t_{e-1} \mbox{ in } \mbox{PM})$.
	\end{itemize}
	Note that our generative process for timestamps in Section \ref{subsec:Time} is sender-oriented where the sender determines when to send the email; thus we incorporate network statistics that depend only on $i$---specifically, the in and outdegrees of sender $i$. 
	
	\subsection{Model selection}\label{subsec:Experiment_email}
	
	The HEM defines a flexible family of models, each of which is defined by a set of features (i.e., the receiver selection features $\boldsymbol{x}$, the selection of event timing features $\boldsymbol{y}$, and the distribution of time increments $f$). Many of these components will be specified based on user expertise (e.g., regarding which features would drive receiver selection), but some decisions may require a data-driven approach to model specification. For example, though theoretical considerations may inform the specification of features, subject-matter expertise is unlikely to inform the decision regarding the family of event timing distribution. Furthermore, since different distribution families (and model specifications more generally) may involve different size parameter spaces, any data-driven approach to model comparison must guard against over-fitting the data. In this section we present a general-purpose approach to evaluating the HEM specification using out-of-sample prediction. We illustrate this approach by comparing alternative distributional families for the event timing component of the model. Here, we specifically compare the predictive performance from two distributions---log-normal and exponential. We particularly choose the log-normal distribution based on some exploratory analysis (e.g., histogram and simple regressions) on raw time increments data, and select the exponential distribution as a baseline alternative that is a commonly specified distribution for time-to-event data, and is also used in the stochastic actor-oriented models (SAOMs) \citep{snijders1996stochastic} as well as their extensions \citep{snijders2007modeling}. 
	\begin{algorithm}[!t]
		\spacingset{1}
		\caption{Out-of-sample predictions}
		\label{alg:PPE}
		\begin{algorithmic}
			\STATE {\bfseries Input}: data $ \{ (s_e, \boldsymbol{r}_e, t_e)\}_{e=1}^E$, 
			number of new data to generate $D$, and initial values of $(\boldsymbol{b}, \boldsymbol{\eta}, \boldsymbol{u}, \sigma^2_\tau)$
			\vskip 0.1in
			\textbf{Test splits}:	
			\STATE draw test senders (out of $E$ senders) 
			\STATE draw test receivers (out of $E\times (A-1)$ receiver indicators $\{\{\boldsymbol{r}_{ej}\}_{j\in [A]_{\backslash s_e}}\}_{e=1}^E$)
			\STATE draw test timestamps  (out of $E$ timestamps) 
			\STATE set the test data as ``missing" (NA)
			\vskip 0.1in
			\textbf{Imputation and inference:}	
			\FOR{$d=1$ {\bfseries to}  $D$}
			\FOR{$e=1$ {\bfseries to}  $E$}
			\IF{$s_e=$  NA}
			\STATE compute $\boldsymbol{\pi}_e$, where $\pi_{ei}=f_{\tau}(\tau_{e}; \mu_{ie}, V(\mu))\times \prod_{i^\prime\neq i}\big(1-F_{\tau}(\tau_{e}; \mu_{i^\prime e}, V(\mu)) \big)$
			\STATE draw $s_e \sim \mbox{Categorical}(\boldsymbol{\pi}_e)$ 
			\ENDIF
			\FOR{$j\in [A]_{\backslash s_e}$}
			\IF{$r_{ej}=$ NA}
			\STATE draw $r_{ej} \sim \mbox{Bernoulli}(p_{ej}),$
			where $p_{ej}=\frac{\exp(\lambda_{iej})}{\exp(\lambda_{iej})+\text{I}(\lVert\boldsymbol{u}_{ie\backslash j}\rVert_1 > 0 )}$ 
			\ENDIF
			\ENDFOR
			\IF{$t_e=$ NA}
			\STATE draw ${\tau}_e$ from its conditional distribution using importance sampling, 
	      where $P({\tau}_e|\, \cdot)\propto f_{\tau}(\tau_{e}; \mu_{s_e e}, V(\mu))\times \prod_{i\neq s_e}\big(1-F_{\tau}(\tau_{e}; \mu_{ie}, V(\mu)) \big)$
			\ENDIF
			\STATE run inference and update $(\boldsymbol{u},\boldsymbol{b}, \boldsymbol{\eta})$ given the imputed and observed data
			\ENDFOR
			\STATE store the estimates for test data
			\ENDFOR
		\end{algorithmic}
	\end{algorithm}
	
	We evaluated the models' ability to predict out-of-sample events and timestamps on Montgomery County email data. We generated a train--test split of the data by randomly selecting 10\% of senders, receivers, and timestamp variables to be held out. Our model then imputed these missing variables during inference, sampling them from their conditional posterior along with the other latent variables. Algorithm \ref{alg:PPE} outlines this procedure in detail. We compare the predictive performance of two versions of our model, each with a different timing distribution over the time increments using $N=500$. We summarize the results of prediction experiments for missing senders, receivers, and timestamps in Figure \ref{figure:PPEresults}. First, we compare the posterior probability of correct senders for each of the missing emails $\{e:s_e=\mbox{NA}\}$, which corresponds to $\pi_{es_e}$ in Algorithm \ref{alg:PPE}. We call this measure the ``correct sender posterior probability." In Figure \ref{subfigure:PPEresults1}, we display boxplots for the distribution of mean correct sender posterior probability---i.e., $\hat{\pi}_{es_{e}} = \frac{1}{N}\sum_{n=1}^N \pi^{(n)}_{es_{e}}$---across the missing emails. The results show that both log-normal and exponential distributions achieve better predictive performance for missing senders compared to what is expected under random guess (i.e., choose one out of $A$ possible senders $=1/18$), with the log-normal model showing better performance than the exponential model. Secondly, since the receiver vector is binary, we compute $F_1$ scores for missing receiver indicators (i.e., all $e$ and $j$ with $r_{ej}$=NA) by taking the harmonic mean of precision and recall:
	\begin{equation}
		\begin{aligned}
			F_1 =2\cdot\frac{\mbox{precision}\cdot \mbox{recall}}{\mbox{precision}+ \mbox{recall}}, &\mbox{ where } \\
			\mbox{recall}  = \frac{\mbox{TP}}{\mbox{TP+FN}} \mbox{ and } \mbox{precision} & =\frac{\mbox{TP}}{\mbox{TP+FP}},
		\end{aligned}
	\end{equation}
with TP denoting true positive (i.e., $\boldsymbol{r}^{\textrm{(obs)}}_{ej}=\boldsymbol{r}^{\textrm{(pred)}}_{ej}=1$), FN denoting false negative (i.e., $\boldsymbol{r}^{\textrm{(obs)}}_{ej}=1$ but $\boldsymbol{r}^{\textrm{(pred)}}_{ej}=0$), and FP denoting false positive (i.e., $\boldsymbol{r}^{\textrm{(obs)}}_{ej}=0$ but $\boldsymbol{r}^{\textrm{(pred)}}_{ej}=1$). Although the generative process for events (Section \ref{subsec: Tie}) is not directly affected by the choice of timestamp distribution, Figure \ref{subfigure:PPEresults2} reveals slight difference between log-normal and exponential in their performance in predicting missing receiver indicators, where log-normal on average outperforms exponential. Finally, we define the prediction error for the $d^{\textrm{th}}$ missing timestamp bot be the median of the absolute relative errors, often referred to as median absolute percentage error (MdAPE), across $N=500$ predictions:
	\begin{figure}[!t]
		\centering
		\begin{tabular}[t]{ccc}
			\begin{subfigure}[b]{0.33\textwidth}
				\caption{Sender prediction}
				\includegraphics[width=\textwidth]{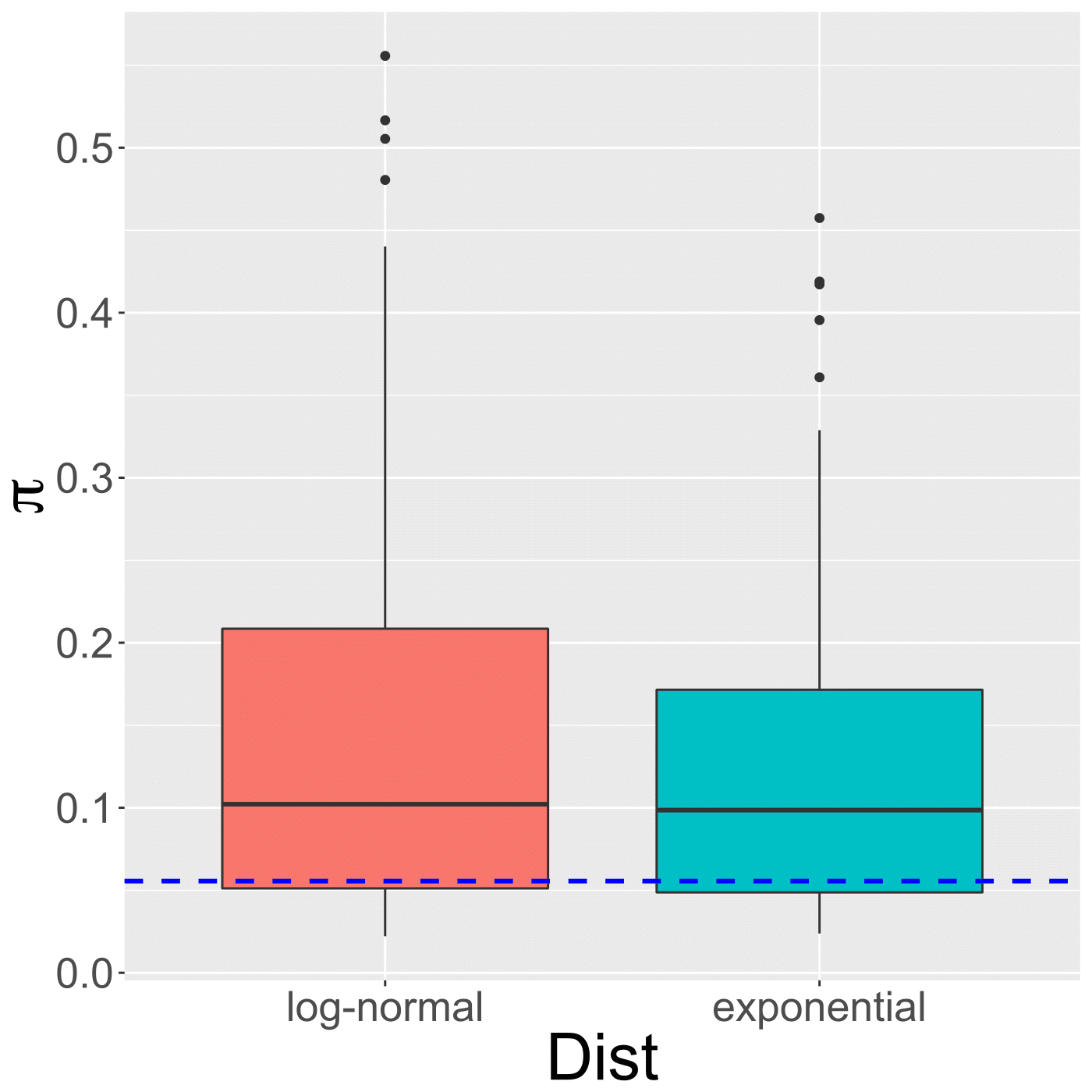}	
				\label{subfigure:PPEresults1}
			\end{subfigure}
			\begin{subfigure}[b]{0.33\textwidth}
				\caption{Receiver prediction}
				\includegraphics[width=\textwidth]{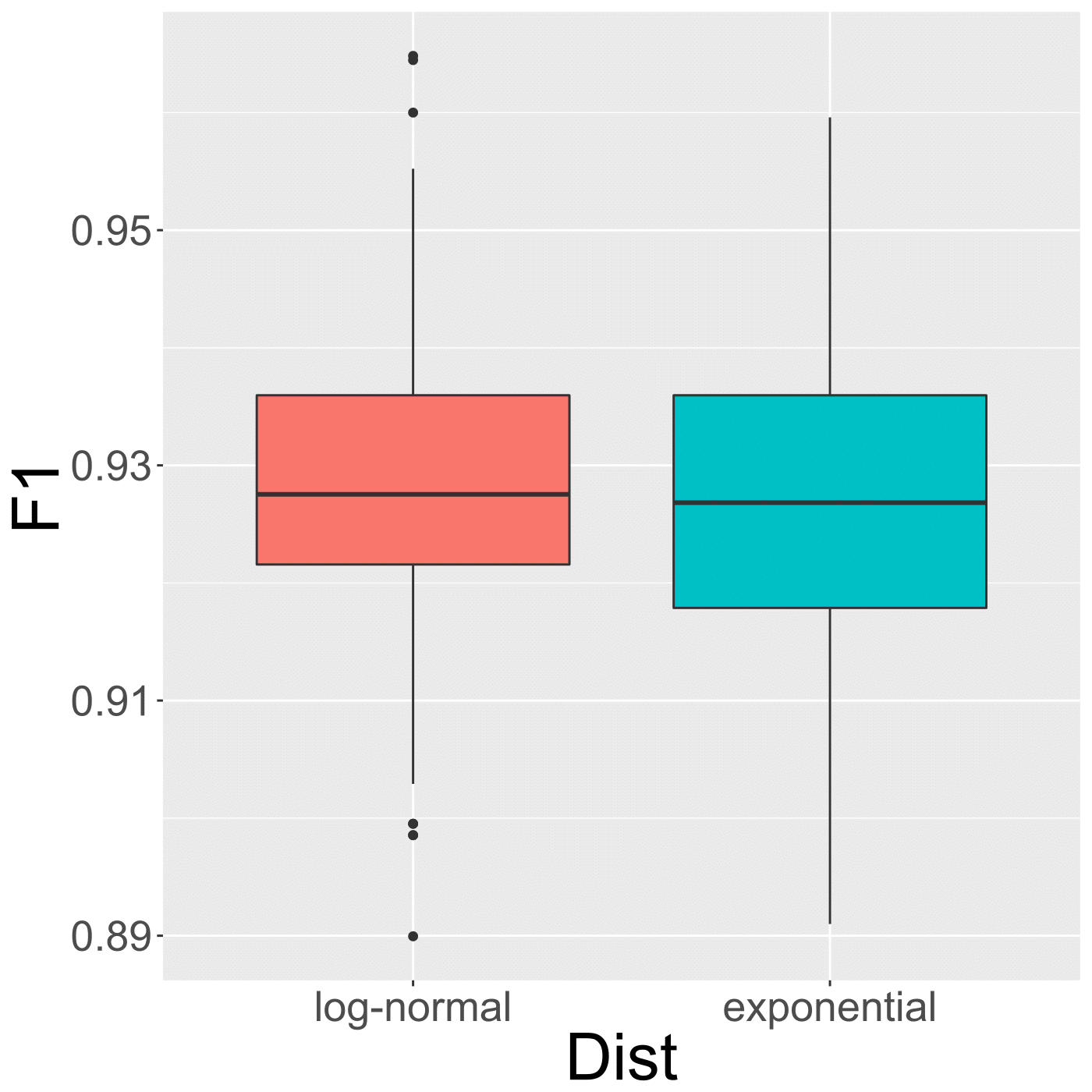}	
				\label{subfigure:PPEresults2}
			\end{subfigure}
			\begin{subfigure}[b]{0.33\textwidth}
				\caption{Timestamp prediction}
				\includegraphics[width=\textwidth]{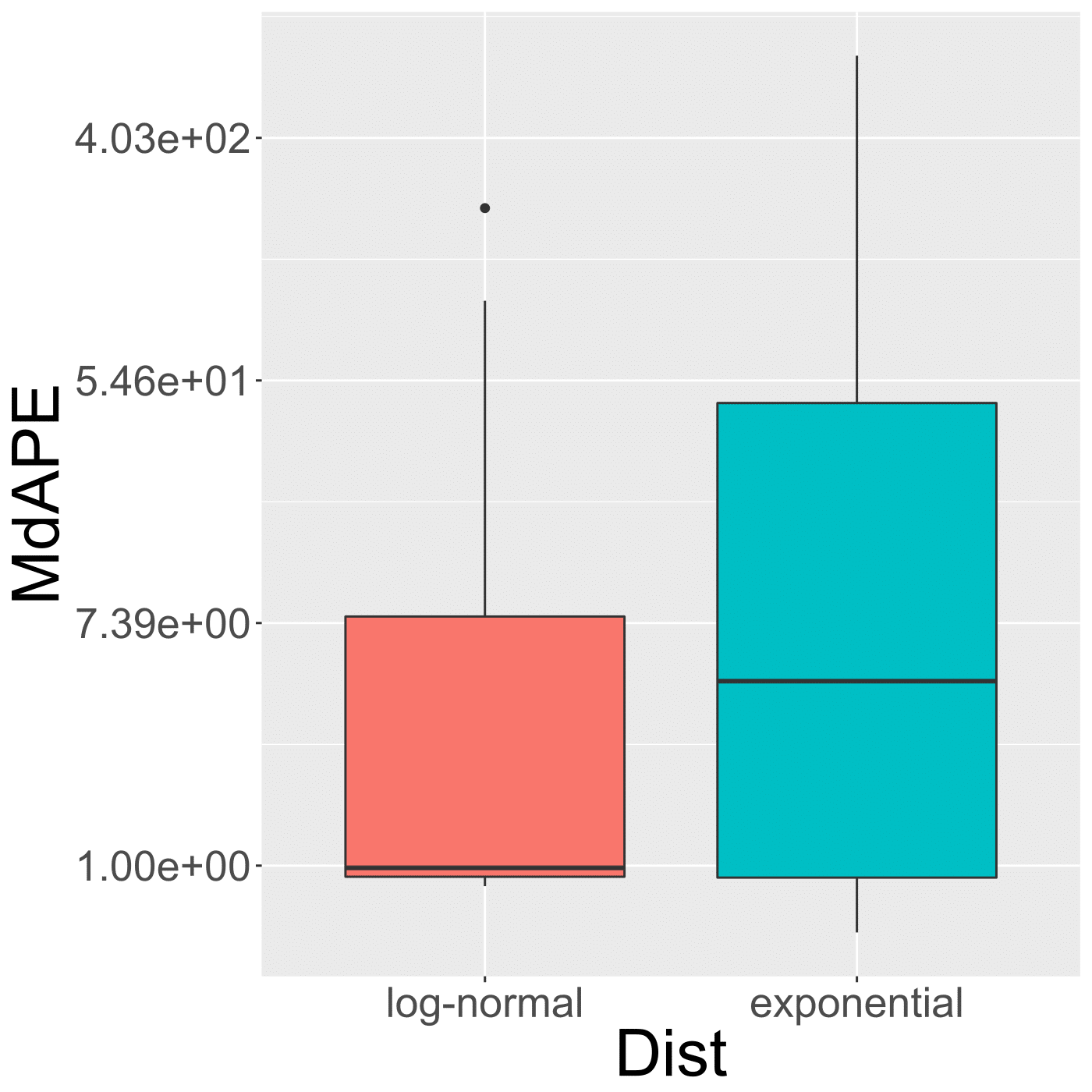}
				\label{subfigure:PPEresults3}	
			\end{subfigure}
		\end{tabular}
		\includegraphics[width=0.4\textwidth]{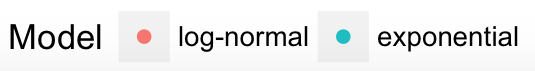}				
		\caption {Comparison of predictive performance between log-normal and exponential distributions: (a) correct sender posterior probability from sender predictions, (b) $F_1$ scores from receiver predictions, and (c) median absolute relative error from timestamp predictions. Blue line in (a) represents the correct sender probability expected by random guess---i.e., $1/A=1/18\approx0.056$.}
		\label{figure:PPEresults}
	\end{figure}		
	\begin{equation}
		\mbox{MdAPE}_{e} = \mbox{median}\Big(\Big\{\abs*{\frac{\tau^{\textrm{(obs)}}_e - \tau^{\textrm{(pred)}_1}_{e}}{\tau^{\textrm{(obs)}}_e}},\ldots, \abs*{\frac{\tau^{\textrm{(obs)}}_e - \tau^{\textrm{(pred)}_N}_{e}}{\tau^{\textrm{(obs)}}_e}}\Big\}\Big).
	\end{equation}
	Figure \ref{subfigure:PPEresults3} presents boxplots for the median absolute percentage errors on a log scale. These plots show that the log-normal distribution fits the time increments significantly better than the exponential distribution. We speculate that this difference can be simply explained by a lack of flexibility in the one-parameter exponential distribution. As illustrated above, we can use this out-of-sample prediction task for two uses---(1) to provide an effective answer to the question ``how does the HEM perform at filling in the missing components of time-stamped network data?" and (2) to offer one standard way to determine the distribution of time increments in Section \ref{subsec:Time}. 
	
	\subsection{Posterior predictive checks}\label{subsec:PPC_email} 	   
	In this section, we perform posterior predictive checks (PPC) \citep{rubin1984bayesianly} to evaluate the appropriateness of our model specification for Montgomery County email data. We formally generated entirely new data by simulating $N=500$ synthetic email datasets $\{(s_{e}, \boldsymbol{r}_{e}, t_{e})\}_{e=1}^E$ from the generative process in Section \ref{sec:generative process}, conditional upon a set of inferred latent variables from inference in Section \ref{subsec:Result_email}. For the test of goodness-of-fit in terms of network dynamics, we use multiple statistics that summarize meaningful aspects of the data: outdegree distribution---the number of emails sent by each node, indegree distribution---the number of emails received by each node, receiver size distribution---the number of receivers on each emails, and a probability--probability (P--P) plot for time increments. 
	\begin{figure}[!t]
		\centering
		\begin{tabular}[t]{cc}
			\begin{subfigure}[b]{0.495\textwidth}
				\caption{Outdegree distribution}
				\includegraphics[width=\textwidth]{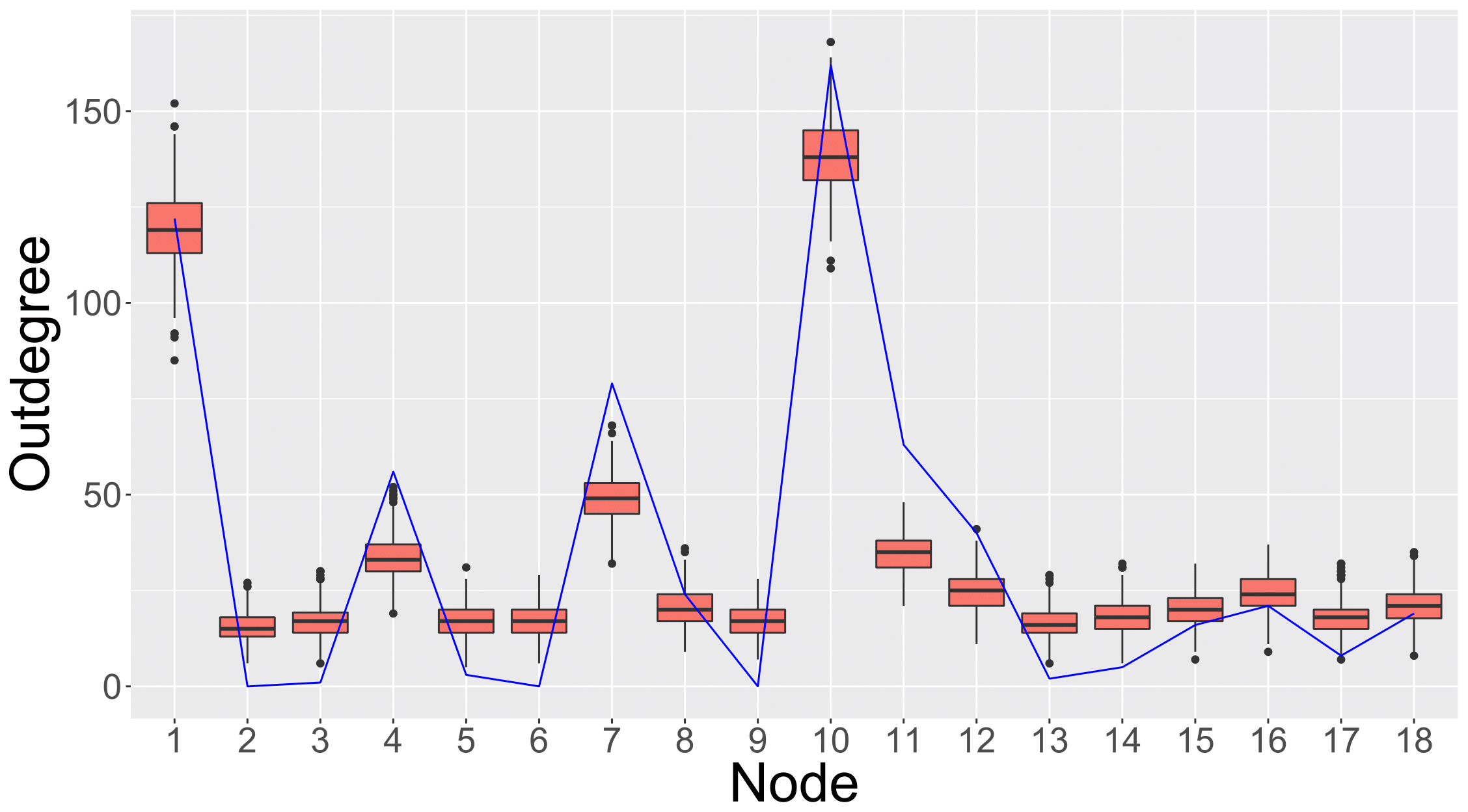}	
			\end{subfigure}
			\begin{subfigure}[b]{0.495\textwidth}
				\caption{Indegree distribution}
				\includegraphics[width=\textwidth]{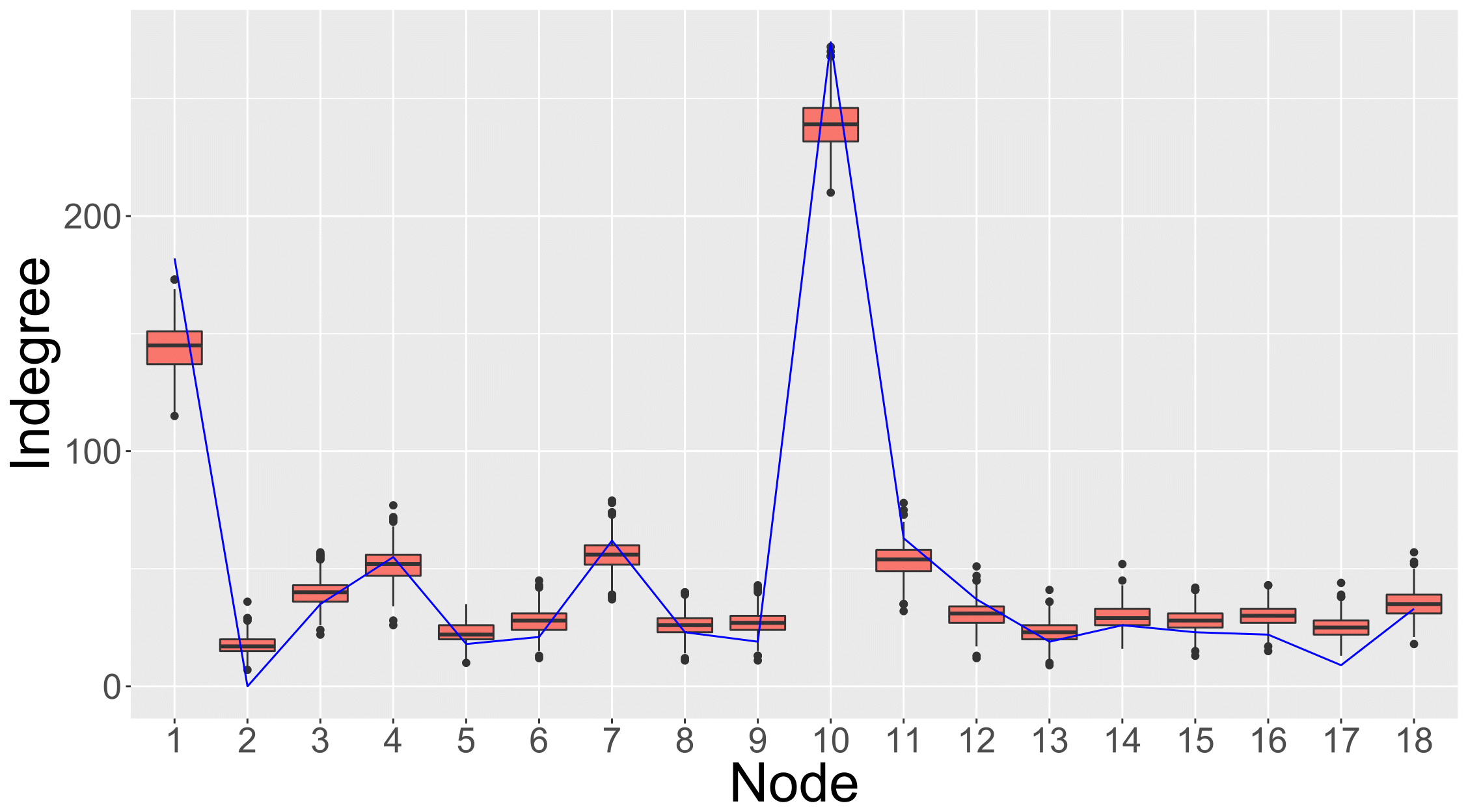}	
			\end{subfigure}\\
			\begin{subfigure}[b]{0.495\textwidth}
				\caption{Receiver size distribution}
				\includegraphics[width=\textwidth]{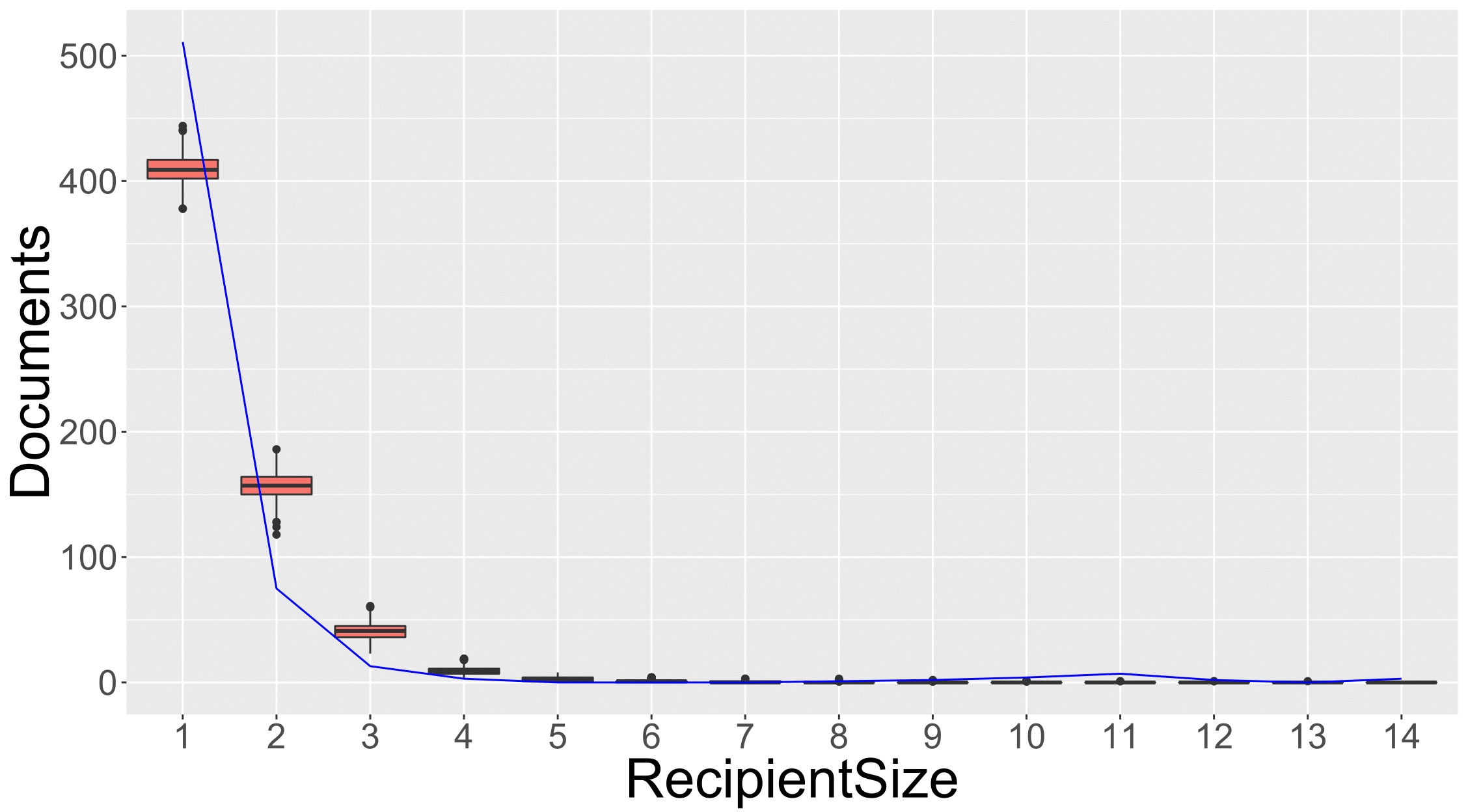}	
			\end{subfigure}
			\begin{subfigure}[b]{0.495\textwidth}
				\centering
				\caption{P--P plot for time increments}
				\includegraphics[width=0.56\textwidth]{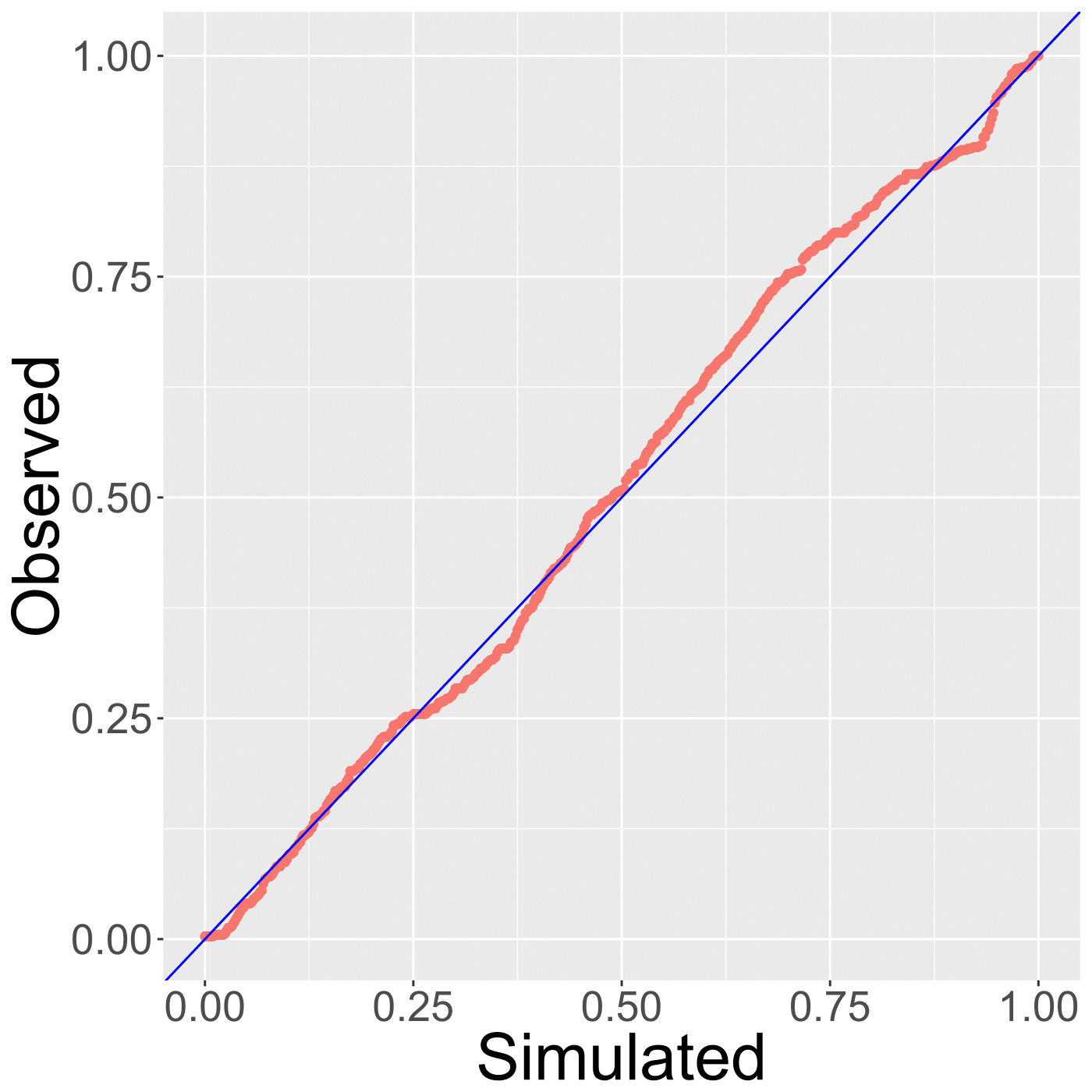}	
			\end{subfigure}
		\end{tabular}
		\caption {PPC results from log-normal distribution. Blue lines denote the observed statistics in (a)--(c) and denotes the diagonal line in (d).}
		\label{figure:PPCresults}
	\end{figure}
	
	Figure \ref{figure:PPCresults} illustrates the results of posterior predictive checks using the log-normal model, which fit the timestamps better than the exponential model (Section \ref{subsec:Experiment_email}). The upper two plots show node-specific posterior predictive degree distributions across $N=500$ synthetic samples, where the left one is for the outdegree statistic and the right plot is for the indegree statistic. For both plots, the x-axis represents the nodes ($a=1,\ldots,18$), and the y-axis represents the number of emails sent or received by the node. When compared with the observed outdegree and indegree statistics (red lines), our model appears to fit the overall distribution of sending and receiving activities across the nodes. For example, node 1 and 10 have a significantly higher level of both sending and receiving activities relative to the rest and this is captured in the model-simulated data. The outdegree distribution of some low-activity nodes are not precisely recovered; however, the  indegree distribution looks much better. Since we use more information in the receiver selection process (i.e., network effects) while we rely solely on minimum time increments when choosing the observed sender, these results are expected. The lower left plot is the distribution of receiver sizes, where the x-axis spans over the size of receivers 1 to 14 (which is the maximum size of observed receivers) and the y-axis denotes the number of emails with x-number of receivers. The result shows that our model is underestimating emails with one receiver while overestimating emails with two, three, and four receivers. One explanation behind what we observe is that the model is trying to recover broadcast emails, which are the emails with $\geq 10$ number of receivers, so that the intercept estimate $b_1$ is slightly moved toward right. It would be an interesting problem in future research to consider how the hyperedge size distribution can be further modified to capture this distribution more accurately. The plot on the lower right is the P--P plot for time increments, which depicts the two cumulative distribution functions---one for simulated time increments and another for observed time increments---against each other in order to assess how closely two data sets agree. Here, the closeness to the diagonal line connecting $(0, 0)$ and $(1, 1)$ gives a measure of difference between the simulated and observed time increments, and our P--P plot shows that we have great performance in reproducing the observed timing distribution. Our findings from the predictive experiments in Section \ref{subsec:Experiment_email} are further revealed in the PPC from exponential distribution, where the PPC plots comparing log-normal and exponential distributions are presented in Appendix C.
	
	\subsection{Exploratory analysis}\label{subsec:Result_email}
	Based on the prediction experiments in Section \ref{subsec:Experiment_email}, we interpret the results from the HEM using the log-normal distribution, with emphasis on understanding the effects of receiver selection and event timing features defined in Section \ref{subsec:Covariates_email}. We assume weakly informative priors for latent variables such as $\boldsymbol{b}\sim N(\boldsymbol{\mu}_b=\boldsymbol{0}, \Sigma_b = 2\times I_P)$, $\boldsymbol{\eta}\sim N(\boldsymbol{\mu}_\eta=\boldsymbol{0}, \Sigma_\eta = 2\times I_Q)$, and $\sigma_\tau^2 \sim \mbox{inverse-Gamma}(a=2, b=1)$, and MCMC (Algorithm \ref{alg:MCMC}) with $O=55,000$ outer iterations and a burn-in of 15,000, where we thin by keeping every $40^{\textrm{th}}$ sample. While the inner iterations for $\sigma_\tau^2$ is fixed as 1, we specify the inner iterations $I_1=20$ for $\boldsymbol{b}$ and $I_2=10$ for $\boldsymbol{\eta}$ to adjust for slower convergence rates. Convergence diagnostics including the traceplots and Geweke diagnostics \citep{geweke1991evaluating} are provided in Appendix D.
	
	\subsubsection{Coefficients for receiver selection features}
\begin{figure}[!t]
				\centering
				\begin{tabular}[t]{cc}
					\begin{subfigure}[b]{0.4975\textwidth}
						\caption{Nodal covariates}
						\includegraphics[width=\textwidth]{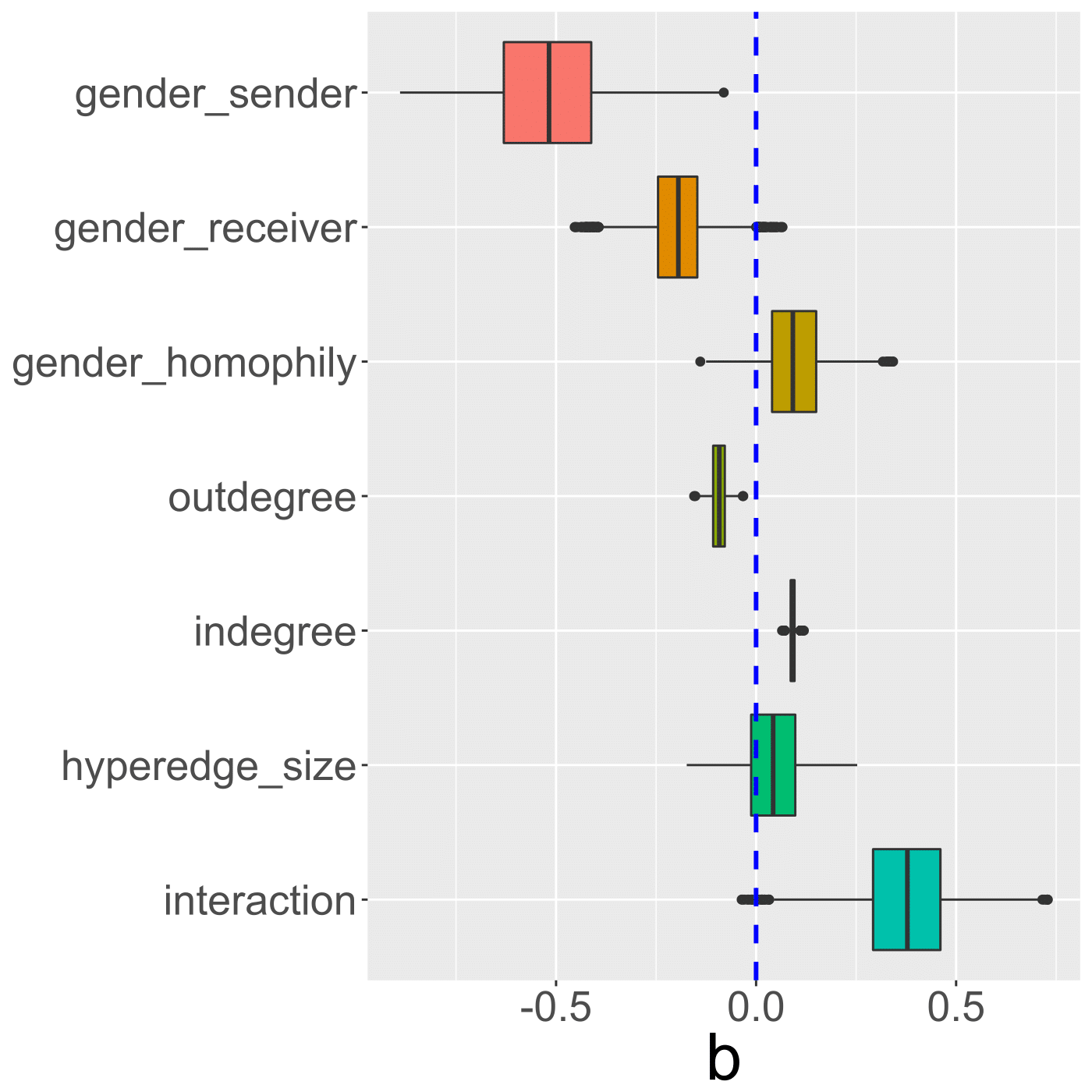}
						\label{subfigure:betaresults1}	
					\end{subfigure}
					\begin{subfigure}[b]{0.4975\textwidth}
						\caption{Dyadic and triadic covariates}
						\includegraphics[width=\textwidth]{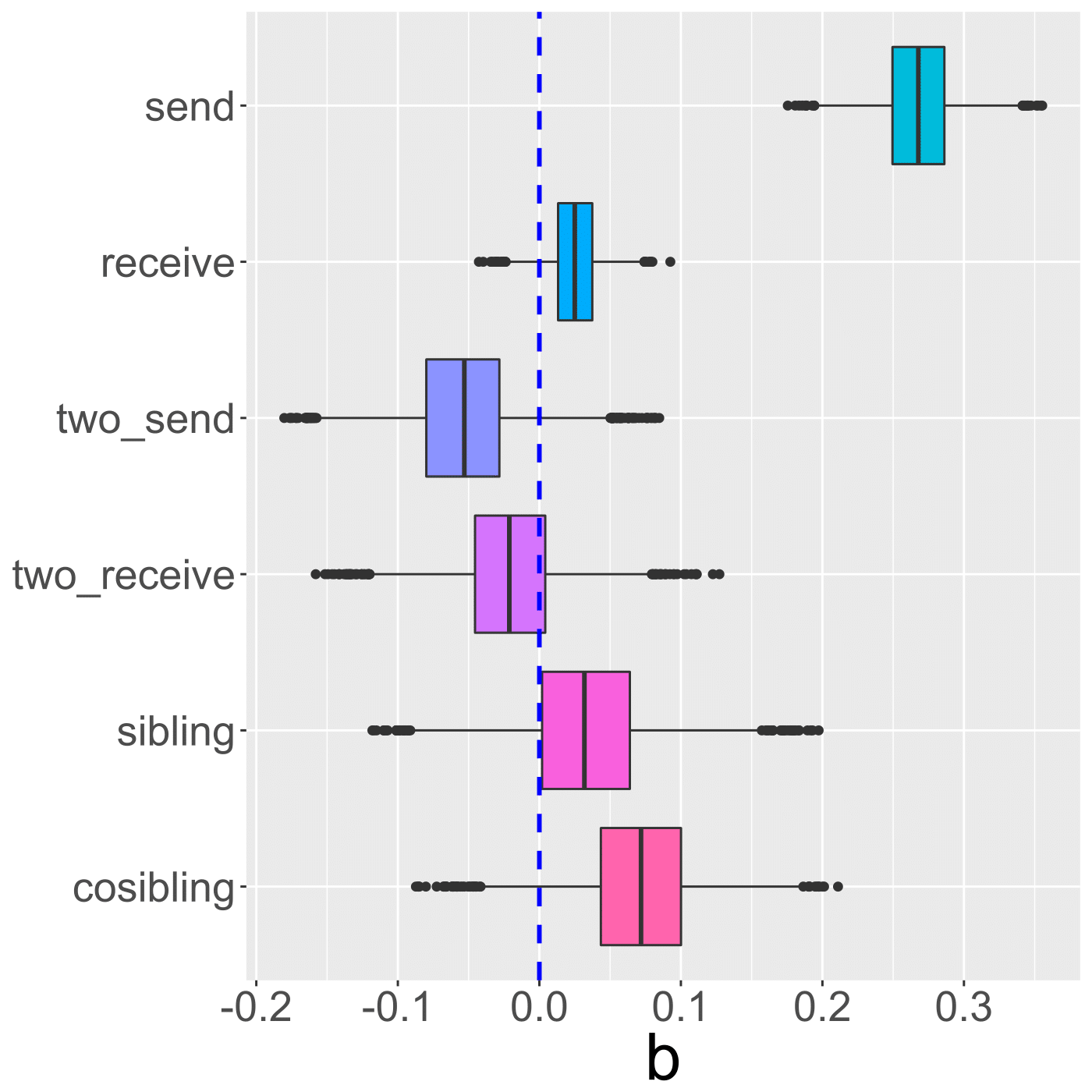}
						\label{subfigure:betaresults2}	
					\end{subfigure}
				\end{tabular}
				\caption {Posterior distribution of $\boldsymbol{b}$ estimates.}
				\label{figure:betaresults}
			\end{figure}

	Figure \ref{figure:betaresults} shows the boxplots summarizing posterior samples of $\boldsymbol{b}$, where Figure \ref{subfigure:betaresults1} displays the coefficients for nodal covariates and \ref{subfigure:betaresults2} displays the coefficients for dyadic and triadic covariates. Since we use the logit functional form 
	\begin{equation*}
		\mbox{logit}(\lambda_{iej})=\log\Big(\frac{\lambda_{iej}}{1-\lambda_{iej}}\Big) =b_{1}+b_{2} x_{iej2}\ldots+b_{14}x_{iej14},
	\end{equation*}
	and can interpret the $\boldsymbol{b}$ estimates in terms of odds ratios $\frac{\lambda_{iej}}{1-\lambda_{iej}}=\exp(b_{1}+b_{2} x_{iej2}\ldots+b_{14}x_{iej14})$. First of all, we find the effects of nodal coavariates ``gender\_\,sender$(i)$" and ``gender\_\,receiver$(j)$" are both nearly always negative in the posterior samples. The log odds that any other node will be added as a receiver of an email is approximately two times less if the sender is a woman. The posterior distribution of the statistic ``outdegree$(i)$" is mostly negative, if sender $i$ sent $n$ number of emails to anyone last week, then sender $i$ is approximately $\exp(-0.109\times n)\approx(0.897)^n$ times less likely to send an email to $j$. However, this straightforward interpretation of the outdegree statistic only applies when the hyperedge size is low. The scenario in which a sender sends a lot of low-hyperedge-size emails may arise due to the use of email for a one-on-one conversation. The large positive estimates of the interaction between hyperedge size and outdegree indicate that those who have recently sent many emails with many receivers on each email are likely to continue sending broadcast emails. This scenario may arise from someone being responsible for distributing timely announcements. When we look at the effect of ``indegree$(j)$," we see a clear popularity effect---those who have received a lot of emails a lot recently are likely to continue receiving a lot of emails. If the receiver $j$ received $n$ number of emails over the last week, sender $i$ is $\exp(0.086\times n)\approx(1.091)^n $ times more likely to send an email to $j$. 
		
	When we look at the effects of dyadic and triadic covariates, one thing that stands out is the large and positive posterior distribution of the statistic ``send$(i,j)$" (i.e., number of times $i$ sent emails to $j$ over the last week) with the posterior mean $\hat{b}_9 = 0.274$, implying that if $i$ sent $n$ number of emails to $j$ last week, then sender $i$ is approximately $\exp(0.274\times n)\approx(1.315)^n$ times more likely to send an email to $j$. The posterior distributions for the reciprocity effect (i.e., ``receive$(i, j)$''), and the four triadic effects, are all fairly evenly spread around zero, so our results do not justify conclusions regarding the nature of these effects in the Montgomery county email network.
	\subsubsection{Coefficients for event timing features}
\begin{figure}[!t]
				\centering
				\includegraphics[width=0.4975\textwidth]{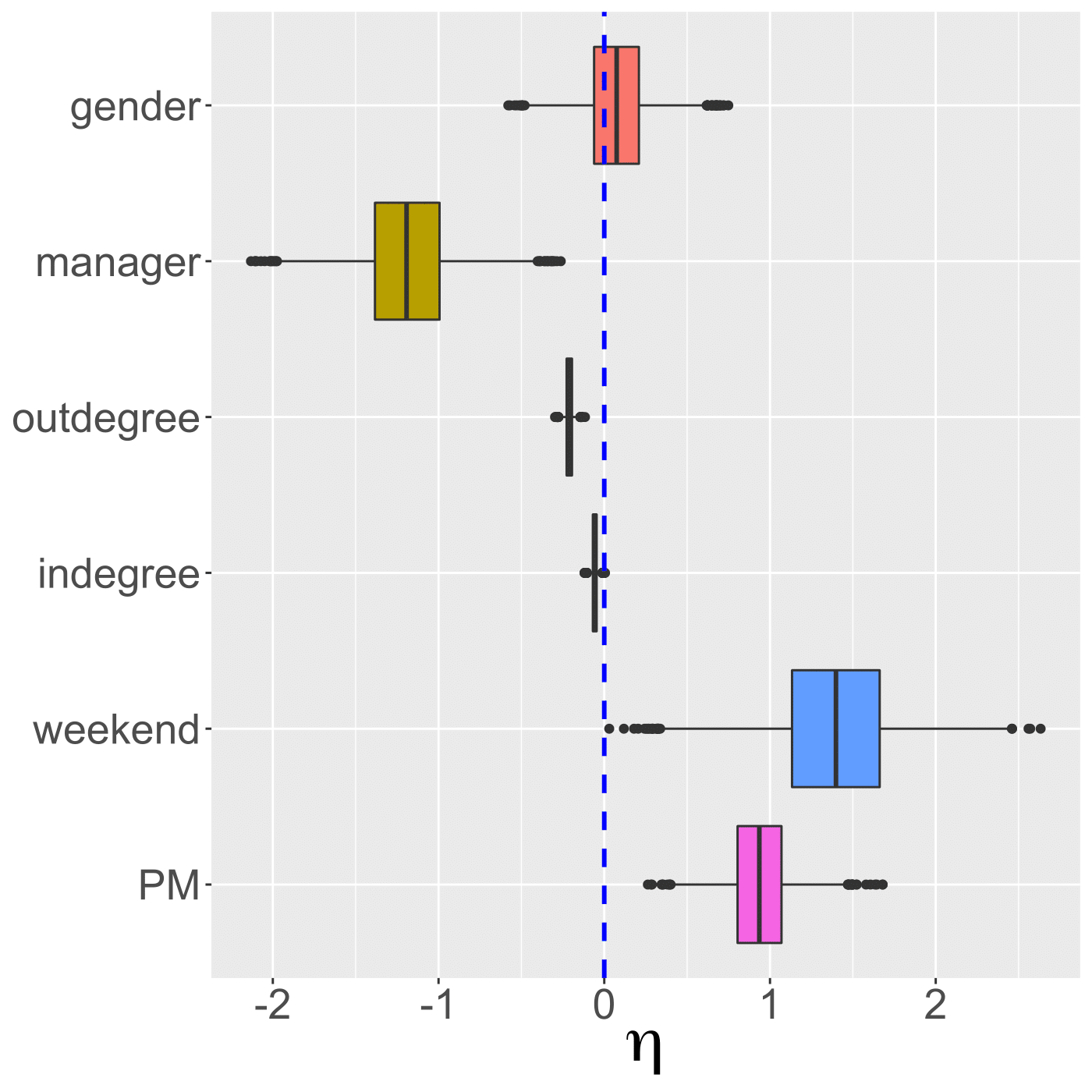}	
				\caption {Posterior distribution of $\boldsymbol{\eta}$ estimates.}
				\label{figure:etaresults}
			\end{figure}

	For event timing features, Figure \ref{figure:etaresults} shows the boxplots summarizing posterior samples of $\boldsymbol{\eta}$. Note that interpretations of the estimated coefficients for $\hat{\boldsymbol{\eta}}$ should be based on the specified time unit of the datset; we specify time units to be hours for the Montgomery county email data. Moreover, since we assume the log-normal distribution for time increments, the coefficients are interpreted in terms of the change in the average log time.
	\begin{equation*}
		\begin{aligned}
			&\log(\tau_{ie}) \sim N(\mu_{ie}, \sigma_\tau^2), \mbox{ with }\\
			&\mu_{ie} = \eta_{1}+\eta_{2} y_{ie2}\ldots+\eta_{7}y_{ie7}.
		\end{aligned}
	\end{equation*}
	The posterior estimates of two temporal effects---``weekend$(e)$" and ``PM$(e)$"---indicate that if the $(e-1)^{\textrm{th}}$ email was sent during the weekend or after midday, then the time to the $e^{\textrm{th}}$ email is expected to take $\exp(1.552)\approx 4.722$ hours and $\exp(0.980)\approx2.665$ hours longer, respectively, compared to their counterparts (i.e., weekdays and am). On the contrary, the covariates ``manager$(i)$", ``outdegree$(i)$", and ``indegree$(i)$"  shorten the amount of time until the next email. For example, being a county manager (i.e., the lead county administrator) lowers the expected value of $\log(\tau_{ie})$ by $\hat{\eta}_3 = -1.070$. The posterior mean estimates for the  ``outdegree$(i)$" and ``indegree$(i)$" statistics are $\hat{\eta}_4=-0.206$ and $\hat{\eta}_5=-0.060$, respectively. These effects indicate that those who are involved in either sending or receiving a lot of emails recently are likely to send emails with greater speed. The posterior distribution for the effect of the gender of the manager is evenly spread around zero. In addition, the posterior mean estimates for the variance parameter $\sigma^2_\tau$ in the log-normal distribution is approximately $\hat{\sigma}^2_\tau=14.093$ with its 95\% credible interval $(12.709, 15.555)$, indicating that there exists large variability in the time increments of emails.
	\section{Conclusion}\label{sec:conclusion}
	Motivated by a growing class of dynamic network models which deal with events recorded in continuous time, the hyperedge event model (HEM) can effectively learn the underlying dynamics in events and their corresponding timestamp formations, providing novel insights to the literature. The HEM explicitly models hyperedges through a receiver selection distribution that forces the sender to select at least one receiver; this obviates the need to preprocess hyperedge data---e.g., by ``duplicating'' hyperedges---to match the assumptions of traditional network models. Our model treating them as pure duplicates. In modeling the timestamps (more precisely time increments) of events, our generalized linear model (GLM) based formulation offers new innovations by eliminating the need to stick with one parameter distribution (e.g., exponential distribution). To our knowledge, the HEM is the only existing model that can be used to generate the sender, receivers, and timestamp of interactions in real time. To make better use of the proposed model, we provide an algorithm for predictive experiments that help to learn which specification of HEM provides a better fit to the data. 
	
	We have demonstrated the effectiveness of our model by analyzing the Montgomery County government emails, where emails serve as a canonical example of directed hyperedge events with one sender and one or more receivers. The estimated effects for receiver selection features reveal that our model is able to understand the structural dynamics similar to those used in the exponential random graph model (ERGM). Our model also learns the effects of event timing features by integrating a survival model for event timing. Although we illustrate the entire framework and application in the context of one type of hyperedge, one sender and one or more receivers, our model can be easily extended to allow the opposite case, one or more sender and one receiver, by slight modification of the generative process (shown in Appendix A). This extension involves promising applications to socio-political networks such as international sanctions and co-sponsorship of bills, and biological networks such as those formed through neural dendrites. 
	
	\begin{acknowledgement}
		This work was supported in part by the University of Massachusetts Amherst Center for Intelligent Information Retrieval and in part by National Science Foundation grants DGE-1144860, SES-1619644, and CISE-1320219. Any opinions, findings, and conclusions or recommendations are those of the authors and do not necessarily reflect those of the
		sponsors.
	\end{acknowledgement}
	
		\newpage
		\bibliographystyle{ba}
			\bibliography{baHEM}	
	\section*{Appendix}
	\subsection*{Appendix A: Alternative generative process} \label{appendix:alternativeGP}
	\begin{algorithm}[H]
		\SetAlgoLined
		\caption{Generative process: one receiver and one or more senders}
		\begin{algorithmic}
			\STATE \textbf{Input}: number of events and nodes $(E, A)$, covariates $(\boldsymbol{x}, \boldsymbol{y})$, and coefficients $(\boldsymbol{b}, \boldsymbol{\eta})$
			\vskip 0.1in				
			\FOR{$e=1$ to $E$}
			\FOR{$j=1$ to $A$}
			\FOR{$i=1$ to $A$ ($i \neq j$)}
			\STATE	set $\lambda_{iej} = {\boldsymbol{b}}^{\top}\boldsymbol{x}_{iej}$
			\ENDFOR
			\STATE	draw $\boldsymbol{u}_{je}  \sim
			\mbox{MB}_G(\boldsymbol{\lambda}_{je})$
			\STATE		set $\mu_{je} = g^{-1}(\boldsymbol{\eta}^\top \boldsymbol{y}_{je})$
			\STATE		draw $\tau_{je} \sim f_\tau(\mu_{je}, V(\mu))$
			\ENDFOR
			\IF {$n \geq 2$ tied events} 
			\STATE	set ${r}_e,\ldots, r_{e+n-1}=\mbox{argmin}_{j}(\tau_{je})$
			\STATE	set $\boldsymbol{s}_e=\boldsymbol{u}_{r_e e},\ldots,\boldsymbol{s}_{e+n-1}=\boldsymbol{u}_{r_{e+n-1} d}$
			\STATE	set $t_e, \ldots, t_{e+n-1}=t_{e-1} + \min_j\tau_{je}$
			\STATE		jump to $e = e+n$
			\ELSE
			\STATE	set ${r}_e = \mbox{argmin}_{j}(\tau_{je}) $
			\STATE	set $\boldsymbol{s}_e= \boldsymbol{u}_{r_e e}$
			\STATE	set $t_e =t_{e-1} + \min_j\tau_{je}$
			\ENDIF
			\ENDFOR
		\end{algorithmic}
		\label{alg:generative2}
	\end{algorithm}
	\subsection*{Appendix B: Normalizing constant of MB$_{G}$}\label{appendix: non-empty Gibbs measure}
	Our probability measure ``MB$_{G}$''---the multivariate Bernoulli distribution with non-empty Gibbs measure---defines the probability of sender $i$ selecting the binary receiver vector $\boldsymbol{u}_{ie}$ as
	\begin{equation*} 
		\begin{aligned}
			& \Pr(\boldsymbol{u}_{ie}|\,\boldsymbol{b}, \boldsymbol{x}_{ie}) = \frac{1}{Z(\boldsymbol{\lambda}_{ie})}\exp\Big(\mbox{log}\big(\text{I}(\lVert \boldsymbol{u}_{ie} \rVert_1 > 0)\big) + \sum_{j \neq i} \lambda_{iej}u_{iej} \Big),
		\end{aligned}
	\end{equation*}
	where the receiver intensity is a linear combination of receiver selection features---i.e., $\lambda_{iej} = {\boldsymbol{b}}^{\top}\boldsymbol{x}_{iej}$---as defined in Secton \ref{subsec: Tie}.
	
	To use this distribution efficiently, we derive a closed-form expression for $Z(\boldsymbol{\lambda}_{ie})$ that does not require brute-force summation over the support of $\boldsymbol{u}_{ie}$ (i.e., $\forall \boldsymbol{u}_{ie} \in [0,1]^A$). We recognize that if $\boldsymbol{u}_{ie}$ were drawn via independent Bernoulli distributions in which $\Pr({u}_{iej}=1|\,\boldsymbol{b}, \boldsymbol{x}_{ie})$ was given by logit$(\lambda_{iej})$, then
	\begin{equation*}
		\Pr(\boldsymbol{u}_{ie}|\,\boldsymbol{b}, \boldsymbol{x}_{ie}) \propto \exp\Big(\sum_{j\neq i} \lambda_{iej}u_{iej}\Big).
	\end{equation*}
	This is straightforward to verify by looking at 
	\begin{equation*}
		\begin{aligned}
			&\Pr(u_{iej}=1|\,\boldsymbol{u}_{ie\backslash j}, \boldsymbol{b}, \boldsymbol{x}_{ie})
			=\frac{\exp{(\lambda_{iej})}}{\exp{(\lambda_{iej})} + 1},
		\end{aligned}\end{equation*}
		where the subscript ``$\backslash j$'' denotes a quantity excluding data from position $j$. Now we denote the logistic-Bernoulli normalizing constant as $Z^{l}(\boldsymbol{\lambda}_{ie})$, which is defined as 
		\begin{equation*}
			Z^{l}( \boldsymbol{\lambda}_{ie})=\sum_{\boldsymbol{u}_{ie} \in [0,1]^{A}} \exp\Big(\sum_{j\neq i} \lambda_{iej}u_{iej}\Big).
		\end{equation*}
		Now, since 
		\begin{equation*}
			\begin{aligned}
				&\exp\Big(\mbox{log}\Big(\text{I}(\lVert \boldsymbol{u}_{ie} \rVert_1 > 0)\Big) + \sum_{j\neq i} \lambda_{iej}u_{iej} \Big)= \exp\Big( \sum_{j \neq i} \lambda_{iej}u_{iej} \Big),
			\end{aligned}
		\end{equation*}
		except when $\lVert \boldsymbol{u}_{ie} \rVert_1=0$, we note that 
		\begin{equation*}
			\begin{aligned}
				Z(\boldsymbol{\lambda}_{ie})& = Z^{l}(\boldsymbol{\lambda}_{ie}) -\exp\Big( \sum\limits_{\forall u_{iej}=0}\lambda_{iej}u_{iej} \Big)
				\\& = Z^{l}(\boldsymbol{\lambda}_{ie}) -  1.
			\end{aligned}
		\end{equation*}
		We can therefore derive a closed form expression for $Z(\boldsymbol{\lambda}_{ie})$ via a closed form expression for $Z^{l}(\boldsymbol{\lambda}_{ie})$. This can be done by looking at the probability of the zero vector under the logistic-Bernoulli model:
		\begin{equation*}
			\begin{aligned}
				&\frac{1}{Z^{l}(\boldsymbol{\lambda}_{ie})}\exp\Big(\sum\limits_{\forall u_{iej}=0}\lambda_{iej}u_{iej} \Big)= \prod_{j \neq i}   \Big(1-\frac{ \exp{(\lambda_{iej})}}{\exp{(\lambda_{iej})} + 1}\Big).
			\end{aligned}  
		\end{equation*}
		Then, we have 
		\begin{equation*}
			\begin{aligned}
				& \frac{1}{Z^{l}(\boldsymbol{\lambda}_{ie})} &= \prod\limits_{j \neq i}\frac{1}{ \exp(\lambda_{iej})+ 1}.
			\end{aligned}  
		\end{equation*}
		Finally, the closed form expression for the normalizing constant is  
		\begin{equation*}
			\begin{aligned}Z(\boldsymbol{\lambda}_{ie}) = \prod_{j \neq i} \big(\mbox{exp}(\lambda_{iej}) + 1\big)-1.
			\end{aligned}  
		\end{equation*}

		\subsection*{Appendix C: Comparison of PPC results: log-normal vs. exponential}\label{appendix: PPCexp}
		\begin{figure}[H]
			\centering
			\begin{tabular}[t]{cc}
				\begin{subfigure}[b]{0.495\textwidth}
					\caption{Outdegree distribution}
					\includegraphics[width=\textwidth]{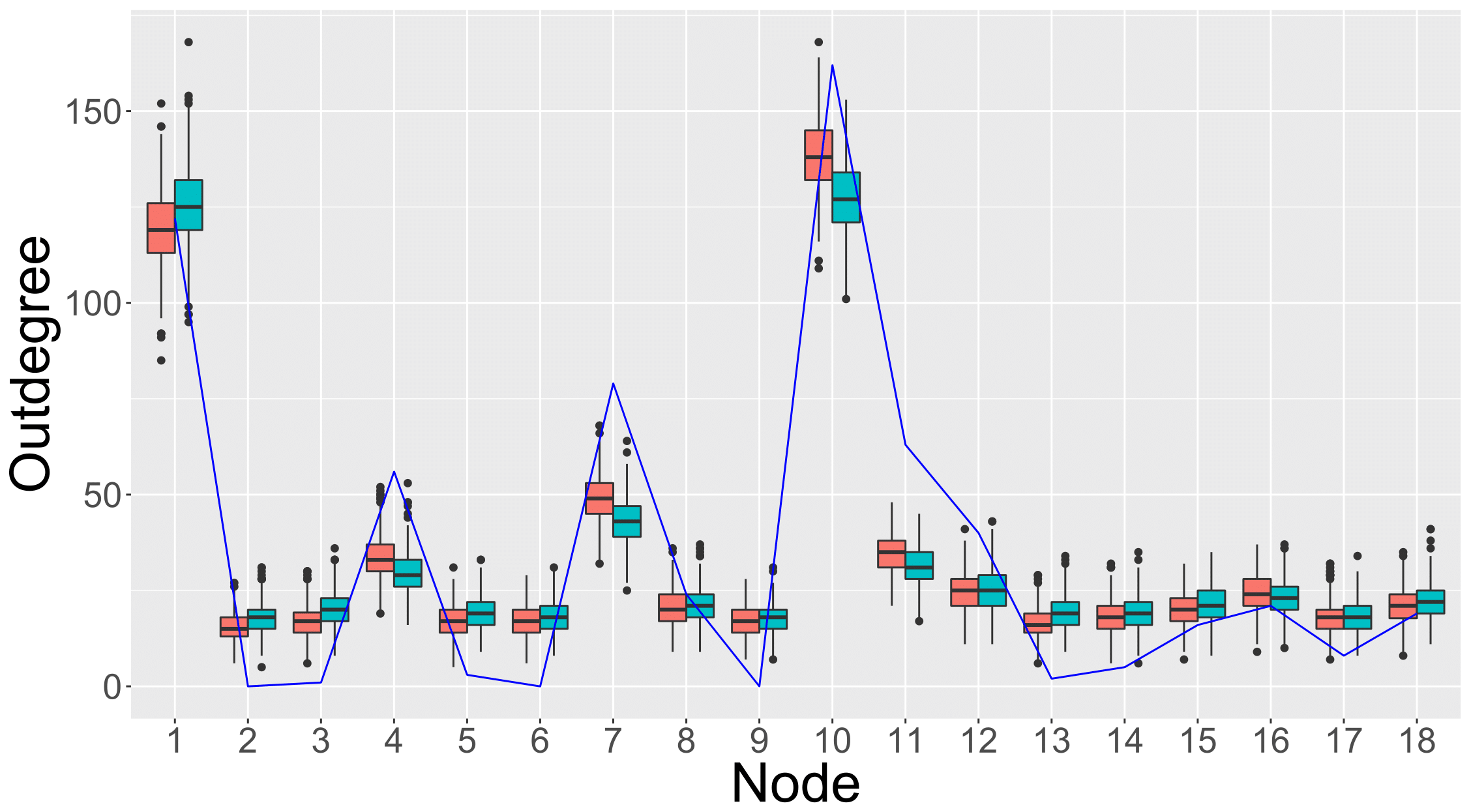}	
				\end{subfigure}
				\begin{subfigure}[b]{0.495\textwidth}
					\caption{Indegree distribution}
					\includegraphics[width=\textwidth]{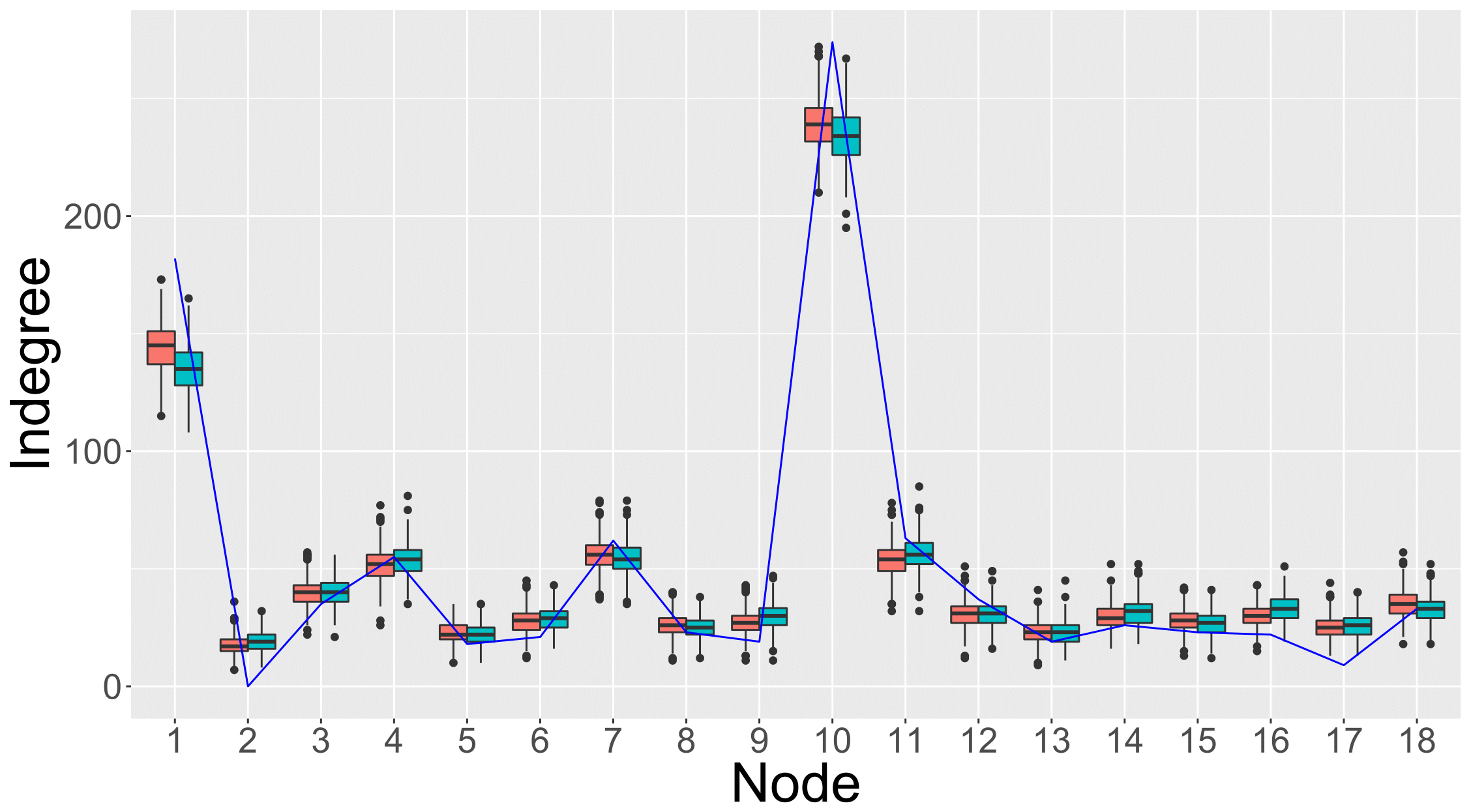}	
				\end{subfigure}\\
				\begin{subfigure}[b]{0.495\textwidth}
					\caption{Receiver size distribution}
					\includegraphics[width=\textwidth]{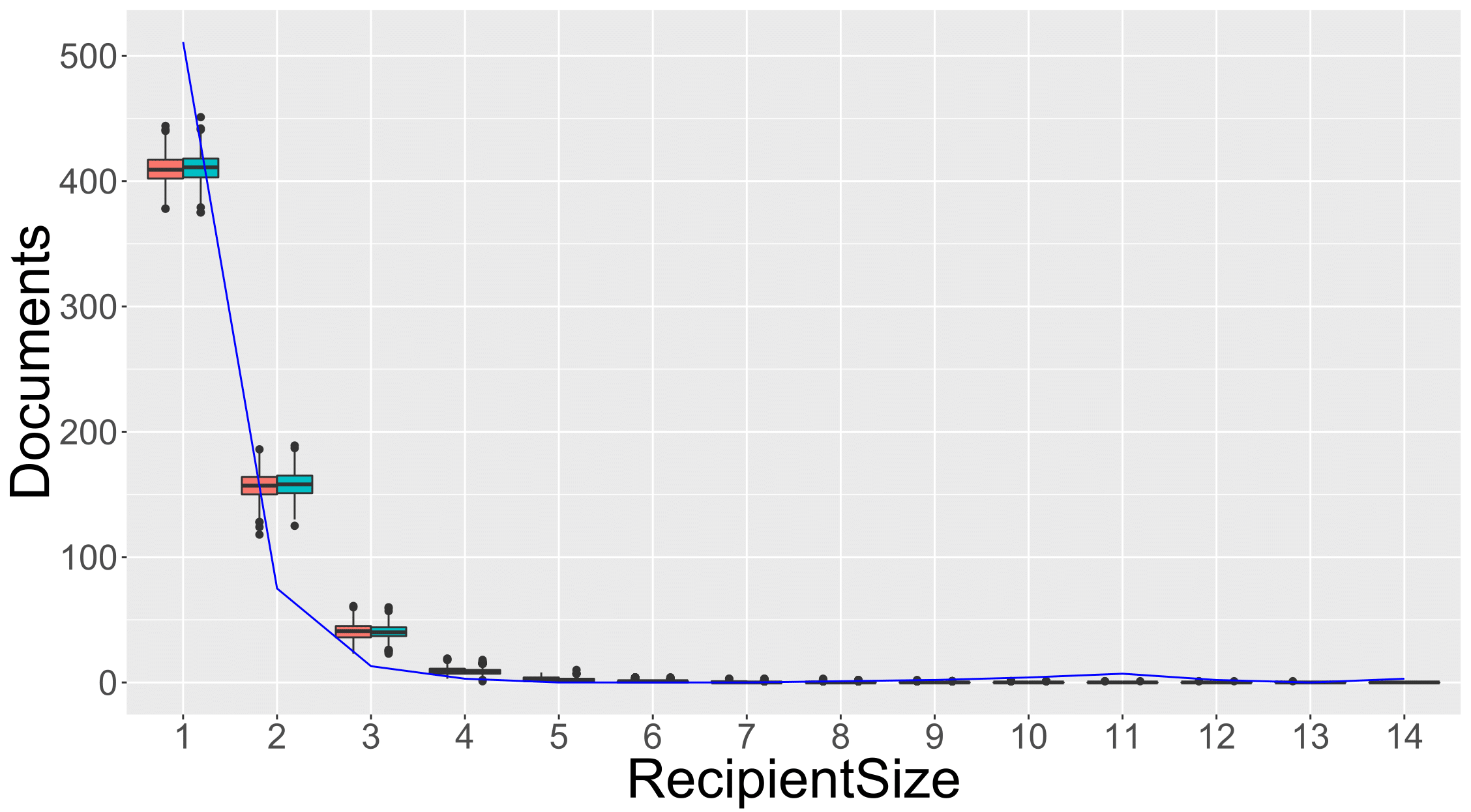}	
				\end{subfigure}
				\begin{subfigure}[b]{0.495\textwidth}
					\centering
					\caption{P--P plot for time increments}
					\includegraphics[width=0.56\textwidth]{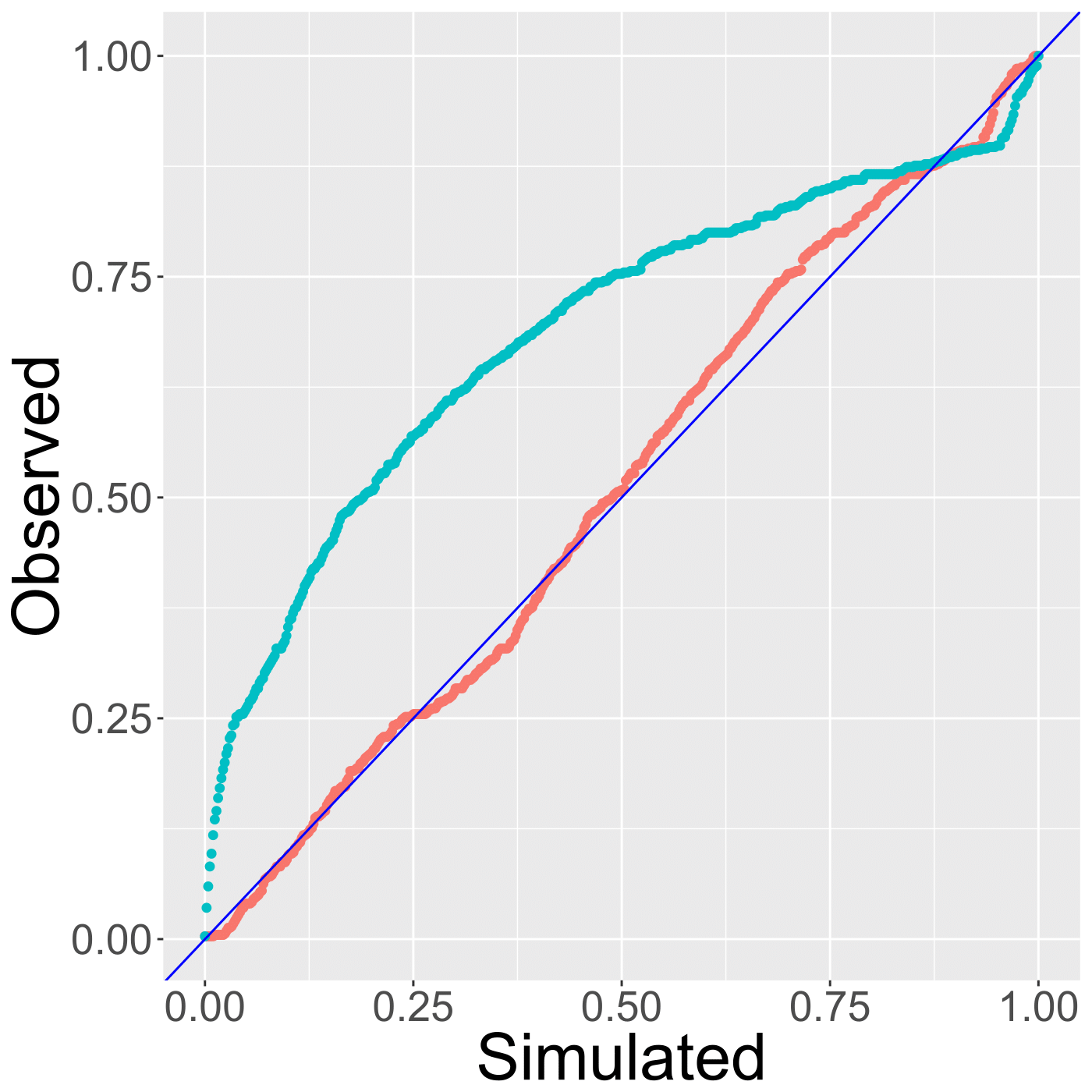}
				\end{subfigure}
			\end{tabular}
			\includegraphics[width=0.4\textwidth]{modellabel.png}
			\caption {Comparison of PPC results between log-normal (\textit{red}) and exponential (\textit{green}) distributions. Blue lines denote the observed statistics in (a)--(c) and denotes the diagonal line in (d).}
			\label{figure:PPCtwo}
		\end{figure}
		\newpage
		\subsection*{Appendix D: Convergence diagnostics}\label{appendix: convergence}
		\begin{figure}[H]
			\centering
			\begin{tabular}[t]{cc}
				\begin{subfigure}[b]{0.495\textwidth}
					\caption{Traceplots of $\boldsymbol{b}$}
					\includegraphics[width=\textwidth]{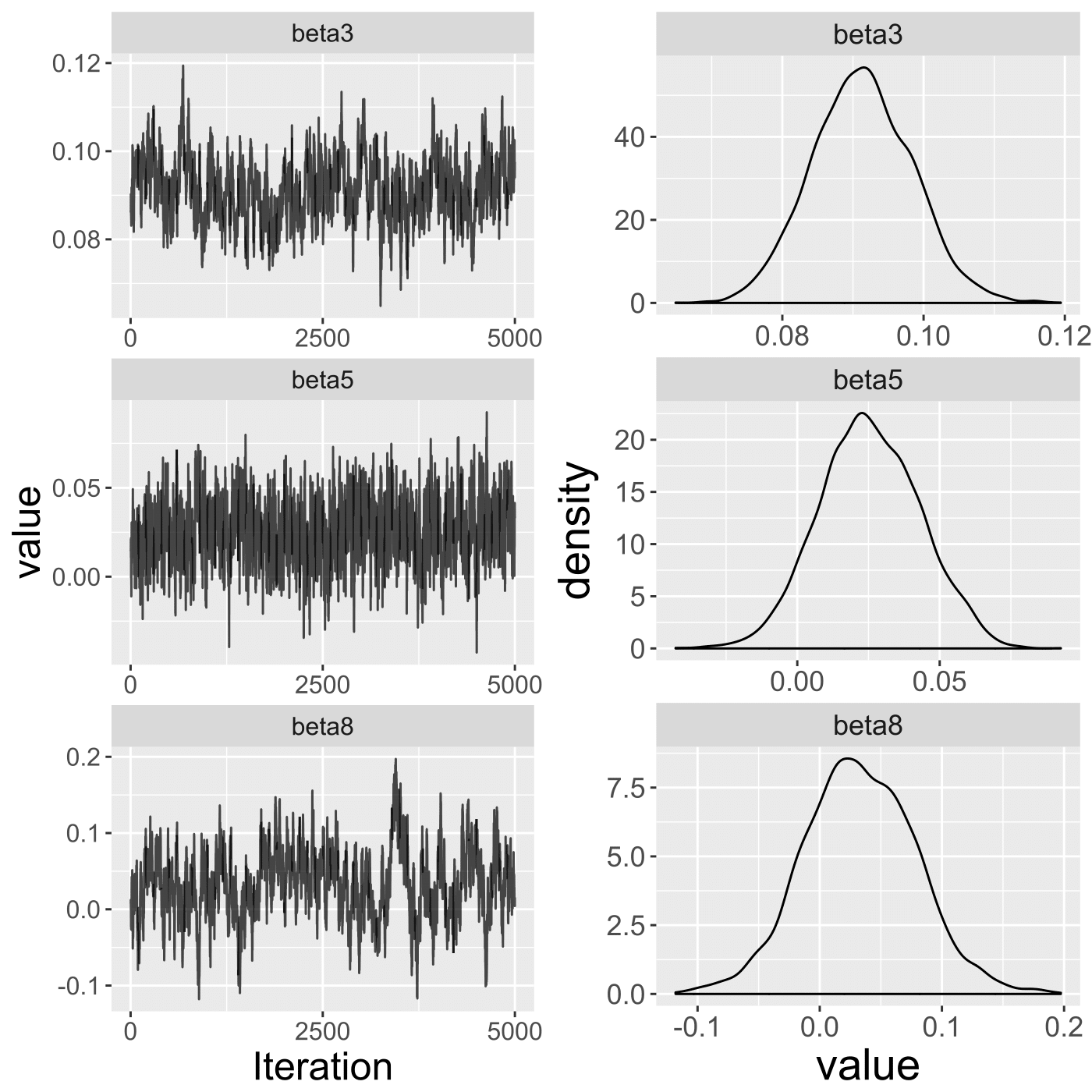}	
				\end{subfigure}
				\begin{subfigure}[b]{0.495\textwidth}
					\caption{Traceplot of $\boldsymbol{\eta}$}
					\includegraphics[width=\textwidth]{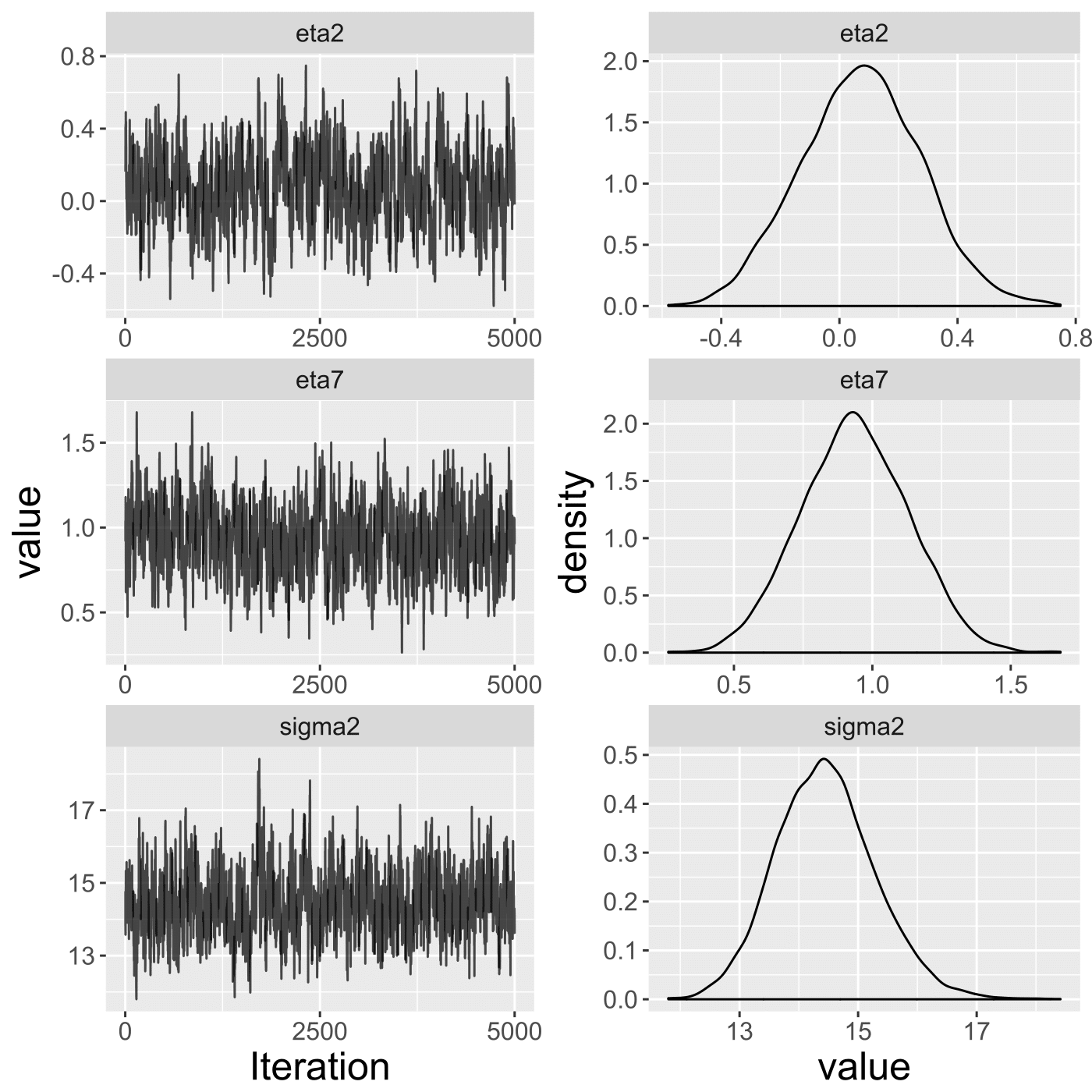}	
				\end{subfigure}\\
				\begin{subfigure}[b]{0.495\textwidth}
					\caption{Geweke diagnostics for $\boldsymbol{b}$}
					\includegraphics[width=\textwidth]{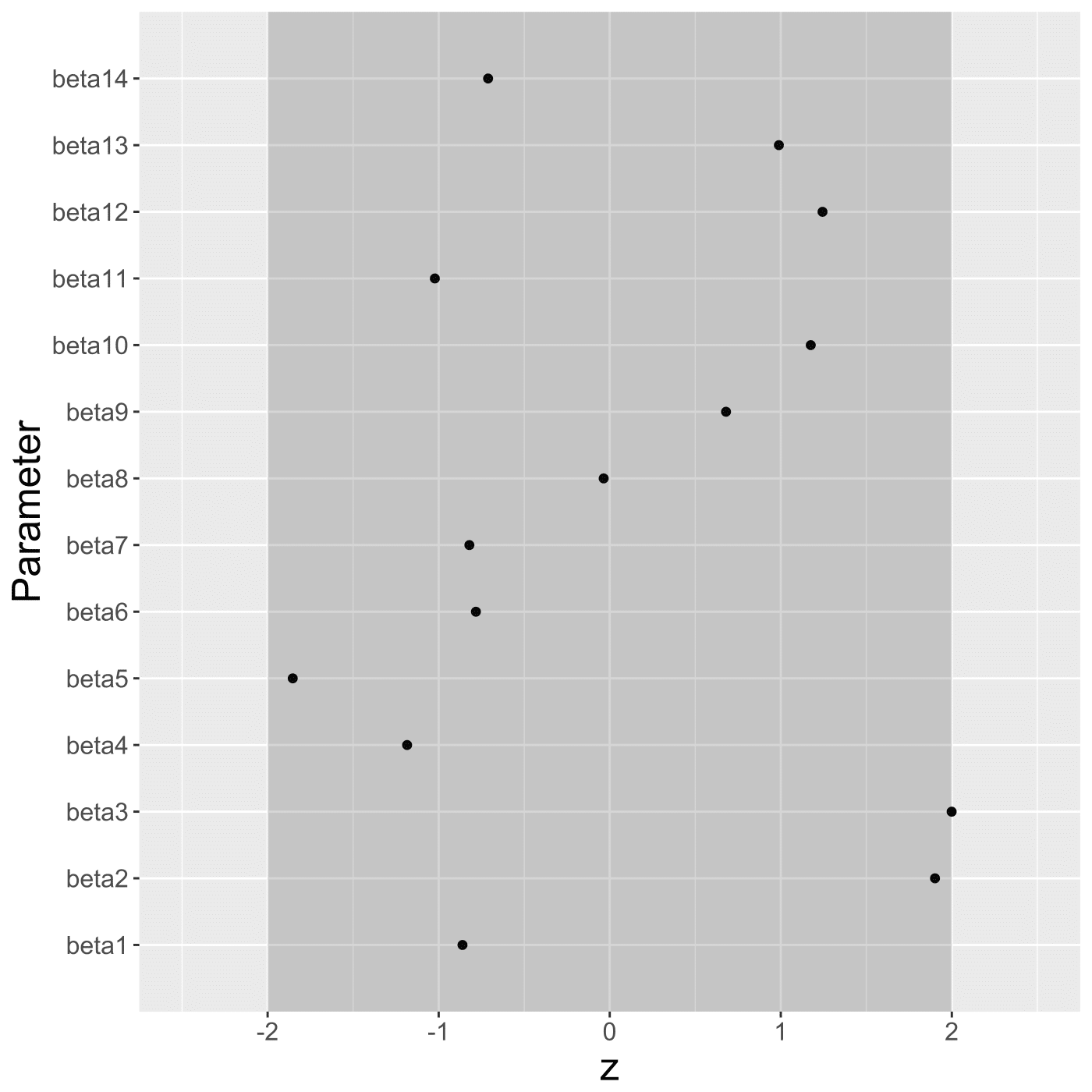}	
				\end{subfigure}
				\begin{subfigure}[b]{0.495\textwidth}
					\centering
					\caption{Geweke diagnostics for $\boldsymbol{\eta}$ and $\sigma^2_\tau$}
					\includegraphics[width=\textwidth]{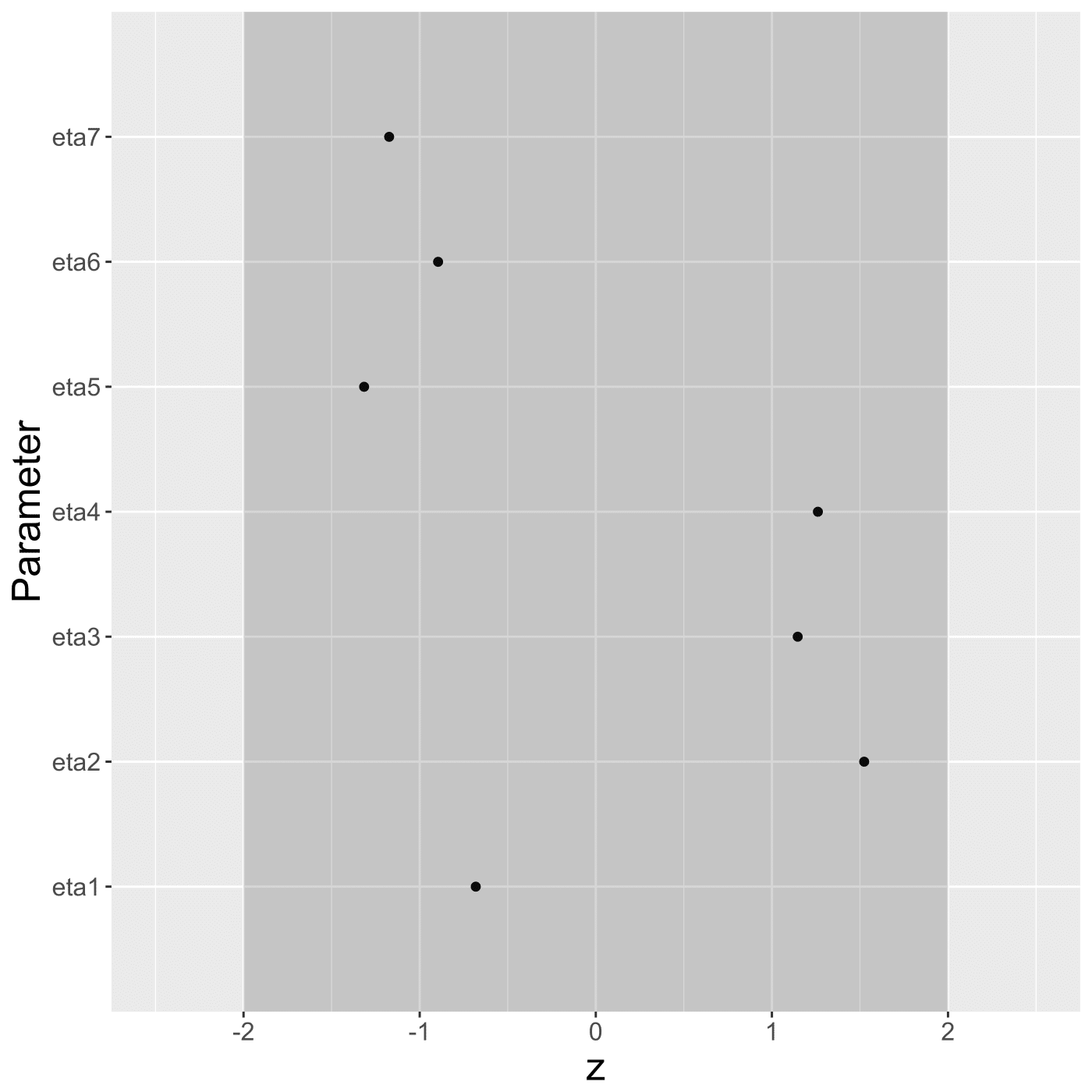}
				\end{subfigure}
			\end{tabular}
			\caption {Convergence diagnostics from log-normal distribution.}
			\label{figure:convergencediag}
		\end{figure}

	\end{document}